\documentclass[aps,prx,twocolumn,superscriptaddress,longbibliography]{revtex4-1}  

\usepackage{graphicx}
\usepackage{amsmath,amssymb}
\usepackage{dsfont}
\usepackage{bm}
\usepackage{color}
\usepackage{MnSymbol, marvosym, wasysym}

\AtBeginDocument{%
    \newwrite\bibnotes
    \def\bibnotesext{Notes.bib}
    \immediate\openout\bibnotes=\jobname\bibnotesext
    \immediate\write\bibnotes{@CONTROL{REVTEX41Control}}
    \immediate\write\bibnotes{@CONTROL{%
    apsrev41Control,author="08",editor="1",pages="1",title="0",year="1"}}
     \if@filesw
     \immediate\write\@auxout{\string\citation{apsrev41Control}}%
    \fi
}%

\renewcommand{\vec}[1]{\boldsymbol{#1}}
\newcommand{\ket}[1]{|#1\rangle}

\newcommand{\up}{{\uparrow}}
\newcommand{\dw}{{\downarrow}}

\newcommand{\pd}{{\phantom{\dagger}}}
\newcommand{\bs}[1]{\boldsymbol{#1}}

\newcommand{\ie}{{\it i.e.},\ }

\usepackage{color}
\definecolor{darkgreen}{rgb}{0.10, 0.65, 0.10}

\newcommand{\sr}[1]{{\color{black}#1}}

\begin{document} 

\title{Majorana modes with side features in magnet-superconductor hybrid systems} 

\author{Daniel Crawford}
\affiliation{School of Physics, University of Melbourne, Parkville, VIC 3010, Australia}
\author{Eric Mascot}
\affiliation{School of Physics, University of Melbourne, Parkville, VIC 3010, Australia}
\affiliation{Department of Physics, University of Illinois at Chicago, Chicago, IL 60607, USA}
\author{Makoto Shimizu}
\affiliation{Department of Physics, Okayama University, Okayama 700-8530, Japan}
\author{Philip Beck}
\affiliation{Department of Physics, University of Hamburg, D-20355 Hamburg, Germany}
\author{Jens Wiebe}
\affiliation{Department of Physics, University of Hamburg, D-20355 Hamburg, Germany}
\author{Roland Wiesendanger}
\affiliation{Department of Physics, University of Hamburg, D-20355 Hamburg, Germany}
\author{Harald O.\ Jeschke}
\affiliation{Research Institute for Interdisciplinary Science, Okayama University, Okayama 700-8530, Japan}
\author{Dirk K.\ Morr}
\affiliation{Department of Physics, University of Illinois at Chicago, Chicago, IL 60607, USA}
\author{Stephan Rachel}
\email{Corresponding author: {\tt stephan.rachel@unimelb.edu.au}}
\affiliation{School of Physics, University of Melbourne, Parkville, VIC 3010, Australia}

\begin{abstract}
Magnet-superconductor hybrid (MSH) systems represent promising platforms to  
host Majorana zero modes (MZMs), the elemental building blocks for 
fault-tolerant quantum computers. Theoretical description of such MSH 
structures is mostly based on simplified models, not accounting for the 
complexity of real materials. Here, based on density functional theory, we 
derive a 
superconducting 80-band model to study an MSH system 
consisting of a magnetic manganese chain on the $s$ wave superconductor 
niobium. For a wide range of values of the superconducting order parameter, 
the system is a topological superconductor, with MZMs exhibiting 
non-universal spatial patterns and a drastic accumulation of spectral weight 
on both sides along the magnetic chain. 
\sr{These side feature states can be explained by an effective 
model which is guided by the {\it ab initio} results.}
Performing scanning tunneling spectroscopy 
experiments on the same system, we observe a spatial structure in the 
low-energy local density of states that is consistent with the theoretical 
findings. Our results open a first-principle approach to the discovery of 
topological superconductors.
\end{abstract}

\date{\today}

\maketitle

%
%
\section{Introduction}

\begin{figure*}[t!]
\centering
\includegraphics[width=\linewidth]{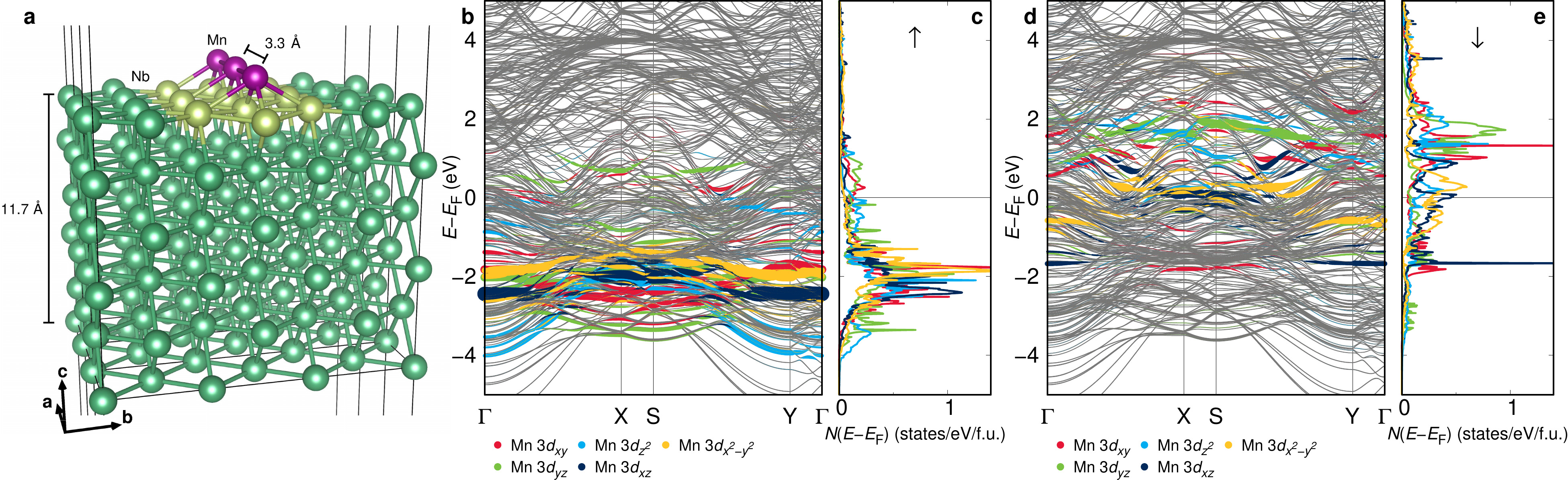}
\caption{\textbf{Ab initio treatment of a Mn chain on the (110) surface of Nb. a,}  Three unit cells used for the DFT calculations are shown; each unit cell contains 36 Nb and one Mn atom. The unit cell is oriented such that $a\,||\,x$, $b\,||\,y$ and $c\,||\,z$, and on
the Nb surface, $a\,||\,[001]$, $b\,||\,[1\bar{1}0]$.
\textbf{b and d,} DFT band structure for spin up and down, respectively, where the orbital character of the Mn $d$-orbitals is highlighted in color while all other orbitals are shown in grey. \textbf{c and e,} Density of states (DOS) for the Mn $d$-orbitals from b and d. Comparison of c and e clearly reveals the spin polarization.}
\label{fig:fig1}
\end{figure*}

Topological superconductors are amongst the most desirable materials, 
as their quasiparticles at zero energy, known as Majorana zero modes (MZMs)\,\cite{kitaev01pu131,mourik-12s1003,albrecht-16n206,deng-12nl6414}, exhibit non-Abelian braiding statistics\,\cite{ivanov01prl268}. These MZMs are promising candidates for topological qubits\,\cite{nayak_non-abelian_2008} and for next-generation quantum computing devices. Magnet-superconductor hybrid (MSH) systems\,\cite{pientka-13prb155420,li-14prb235433,nadj-perge-13prb020407,schecter-16prb140503,klinovaja-13prl186805,martin-12prb144505} offer a particularly appealing platform: (i) they are comprised of well-understood components, magnetic atoms such as Mn, Fe or Co deposited on the surface of a conventional superconductor such as Nb, Re or Pb; (ii) by virtue of single-atom manipulation techniques the magnetic structures can be assembled on the atomic level; (iii) scanning tunneling methods allow the local measurement and manipulation of such structures. Despite the tremendous progress, both experimentally and theoretically, in the past years, one-dimensional MSH structures\,\cite{nadj-perge-14s602,ruby_end_2015,pawlak_probing_2016,jeon-17s772,ruby-17nl4473,kim_toward_2018,schneider-20nc4707,schneider-21np}, {\it i.e.}, magnetic chains on the surface of superconductors with spin-orbit coupling due to the inversion-symmetry breaking at that surface, still struggle with either sample quality or the size of the spectral gap necessary to separate the MZMs from the low-lying excited states. 
On the theoretical side, most simulations are (i) based on simplified 
models\,\cite{choy_majorana_2011,nadj-perge-13prb020407,pientka-13prb155420,crawford-20prb174510,mier-21prb245415} and (ii) often done for entirely one-dimensional structures\,\cite{choy_majorana_2011,pientka-13prb155420}, thus ignoring the complexity and the two- or three-dimensional character of the materials involved. 
In addition, such models often couple both the magnetic Zeeman field (accounting for the magnetic moments) and the superconducting order parameter (accounting for the superconducting proximity effect) to the same orbitals, which is unphysical. Despite these simplifications, such models clearly show topological phases with localized MZMs at chain ends. Early experiments for Fe chains on Pb(110)\,\cite{nadj-perge-14s602} seemed to be in good agreement with such modeling. Subsequent experiments on Fe/Pb(110) revealed a peculiar {\it double eye} feature\,\cite{feldman-17np286}, 
which was explained by the authors of Ref.\,\cite{feldman-17np286} through the interplay of a local suppression of the order parameter in the vicinity of the magnetic chain and the tunneling geometry including the tip trajectory.
However, the aforementioned simplified models in the literature cannot capture features such as the double eye.
In the experimental part of this paper, we present $dI/dV$ scanning tunneling spectroscopy data for Mn/Nb(110) and Fe/Nb(110) systems where characteristic low-energy states are present on the sides of the magnetic chain. 
We perform {\it ab initio} modeling of Mn chains along [001] on Nb(110). We construct an effective superconducting 80-band Bogoliubov-de Gennes model, 
which turns out to be topologically non-trivial for most of the considered parameter range of the superconducting order parameter. 
\sr{Analysis of this model guides us to construct a further simplified model--still containing a four-atomic unit cell and a non-trivial amount of hoppings beyond nearest-neighbors--where the zero-energy states for the topological phase exhibit side-features. 
Further investigation of this model} 
reveals that the spatial structure of the emerging MZMs sensitively depends on the size of the spectral gap, with low-energy features varying between localized along the side of the magnetic chains and localized near the end, with the latter resembling the previously observed double-eye feature.
Due to the promising agreement of the modeling with our scanning tunneling experiments for Mn chains on Nb(110) and the similarities to other MSH structures such as Fe chains on Nb(110) or Pb(110),  
our work questions the notion of point-like MZMs as the hallmark of one-dimensional topological superconductors. 
We furthermore propose that MZMs appearing as side features might be a common property of topologically non-trivial MSH structures in the small gap regime due to the complex interplay of the MZM's orbital composition and their repulsion from the magnetic chain.

%
%
\section{Results}

\begin{figure*}[t!]
\centering
\includegraphics[width=\linewidth]{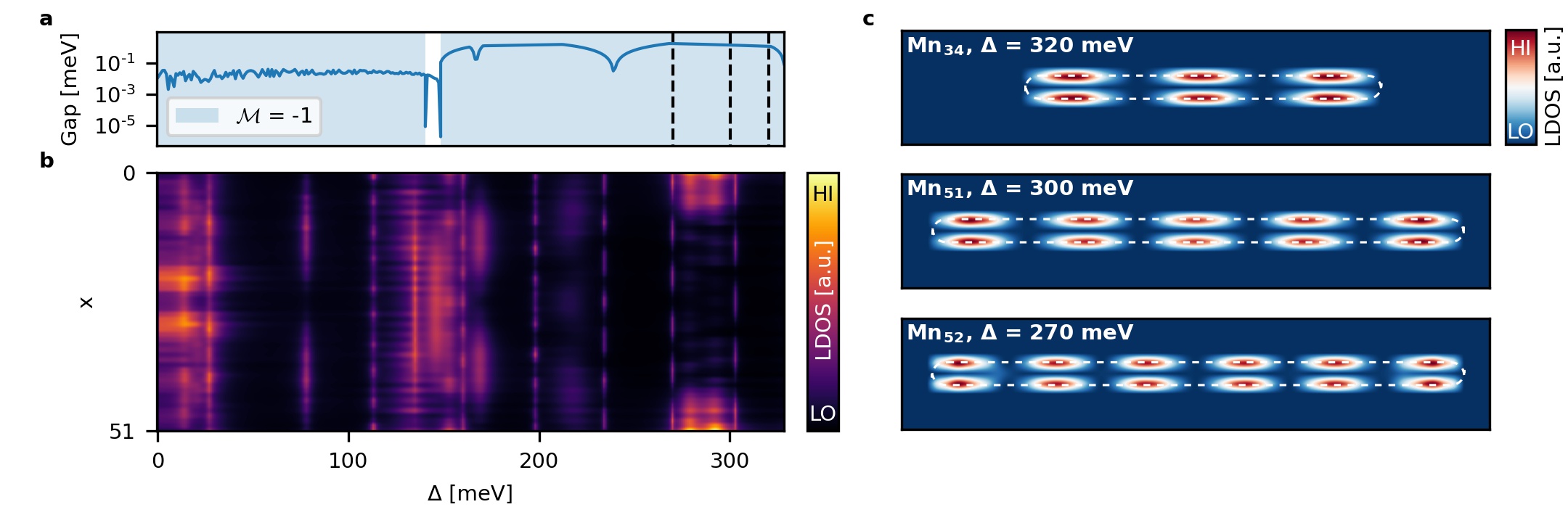}
\caption{\textbf{Analysis of DFT-based superconducting 80-band model. a,}  Topological phase diagram as a function of the bare superconducting amplitude $\Delta$ (topological region in blue with $\mathcal{M}=-1$, and white is trivial). The gap size is shown as the blue curve. \textbf{b,} Zero-energy LDOS as a function of lattice site along $x$ and $\Delta$ (the $y$-dependence has been summed up such that the LDOS appears one-dimensional) for chain length $N=51$. \textbf{c,} Examples of zero-energy LDOS$(x,y)$ for various values of $\Delta$, as indicated by the dashed vertical lines in a, with pronounced side features. Chain length and $\Delta$ are given within the panels.
The theoretical LDOS is always evaluated in the plane of the chain (for a discussion of the three-dimensional LDOS see SI section S4).}
\label{fig:fig2}
\end{figure*}

\textbf{DFT modelling}.~
Using density functional theory (DFT) and projective Wannier functions, we study the electronic structure of Mn chains on Nb(110), and map it onto a tight-binding model to extract the orbital dependence, spin splitting and spin-orbit coupling. To this end, we
construct a slab supercell for the Mn chains along the [001] direction with $d_{\rm Mn-Mn}=3.30$\,{\AA} on Nb(110). 
There is little previous work on {\it ab initio} based modeling of MSH structures, see\,\cite{kobialka-20prb205143,laszloffy-21rXiv2106.09977} for two notable exceptions.
Fig.\,\ref{fig:fig1}a shows the largest supercell that we can study computationally.
The distortions of the relaxed Nb surface with Mn in the optimal hollow
position\,\cite{method} entail that the Nb(110) electronic structure with a Mn chain attached differs from the bulk band structure both because of surface relaxation and because of distortions due to the Mn chain. Fig.\,\ref{fig:fig1}b and d shows the fully relativistic (GGA+SO) spin-polarized band structure of our system, with Mn orbital character highlighted. 
We find that the ferromagnetic arrangement is $29.4$\,meV/Mn lower in energy than the antiferromagnetic one, indicating that Fig.\,\ref{fig:fig1}b and d represents the DFT electronic ground state of the system. 
We find that an out-of-plane alignment of the magnetic moments is slightly lower in energy (by about 0.1\,meV/Mn) compared to an in-plane alignment.
To construct a tight-binding model of the bandstructure  
obtained from the DFT calculations, we use projective Wannier 
functions within FPLO\,\cite{Eschrig2009}. However, the inclusion of all relevant orbitals, the Mn 3$d$, 4$s$ and 4$p$ as well
as all Nb 4$d$, 5$s$ and 5$d$ orbitals, would result in a 666 band tight-binding model, whose investigation, 
for example in terms of the topological invariant or its real-space properties, is computationally beyond our current abilities. 
To arrive at a tractable tight-binding model, we therefore restrict our attention to the Mn 3d orbitals and the
 Nb 4d orbitals that bond with them (see highlighted Nb sites in Fig.\,\ref{fig:fig1}a). This results in a 40-band 
 tight binding model, which (i) correctly reproduces the spectral weight of the Mn $3d$ bands (as given by the DFT results Fig.\,\ref{fig:fig1}\,c and e), 
which are relevant for the emergence of topological superconductivity due to their magnetic moments, and (ii) contains 
the direction dependent $d$ orbital overlaps
as well as the effects of Mn spin splitting and Nb spin-orbit coupling.

%
%

\vspace{20pt}
\noindent
\textbf{80-band model}.~
We convert the DFT-based 40-band tight-binding model into a superconducting 80-band Bo\-go\-liu\-bov--de Gennes model with the {\it bare} superconducting order parameter $\Delta$ as the only free parameter\,\cite{method}.
Note that we apply the superconducting onsite pairing term only to the orbitals of the three Nb atoms but not to the Mn atom; we assume that superconductivity is proximitized into the Mn orbitals, thus accurately describing the experimental system. It is important to emphasize that $\Delta$ here only serves as a phenomenological parameter and does not have a one-to-one correspondence with the superconducting gap (which is known to be $\Delta_{\rm Nb}$=1.51\,meV for Nb). In fact, it is quite clear that the bare $\Delta$ must be chosen to be much larger since the three Nb atoms represent the superconducting proximity effect of the entire (three-dimensional) Nb substrate. In order to 
investigate the topological phase diagram of this model
we compute the gap size and the topological invariant $\mathcal{M}$\,\cite{kitaev01pu131} characteristic of topological class D\,\cite{kitaev09,schnyder-08prb195125}. The system is gapped, but the spectral gap is tiny for $\Delta<130$\,meV (Fig.\,\ref{fig:fig2}a) 
\sr{and ranges between $7$\,$\mu$eV and $188$\,$\mu$eV for periodic boundary conditions (mean value $69$\,$\mu$eV).
}
Moreover, we find for $0 < \Delta <120$\,meV and for $\Delta >130$\,meV the ground state to be topologically non-trivial, {\it i.e.}, $\mathcal{M}=-1$ (Fig.\,\ref{fig:fig2}\,a). 
\sr{Gap sizes for the regime $\Delta >130$\,meV are in the range from $33$\,$\mu$eV to $1.83$\,meV (mean value $1.19$\,meV).}
Only the narrow region $120$\,meV $<\Delta<$ 130\,meV turns out to be trivial. Our momentum space analysis thus reveals that 
Mn chains in the [001] direction on Nb(110) realize a topological superconductor, which due to bulk-boundary correspondence is accompanied by MZMs that are localized 
at the chain ends. Chain length and effective gap size might, however, influence MZMs and cause them to hybridize in an oscillatory fashion as a function of length, which is one of the key signatures of hybridized MZMs in short chains\,\cite{stanescu-13prb094518,schneider2}.

\begin{figure*}[t!]
\centering
\includegraphics[width=\linewidth]{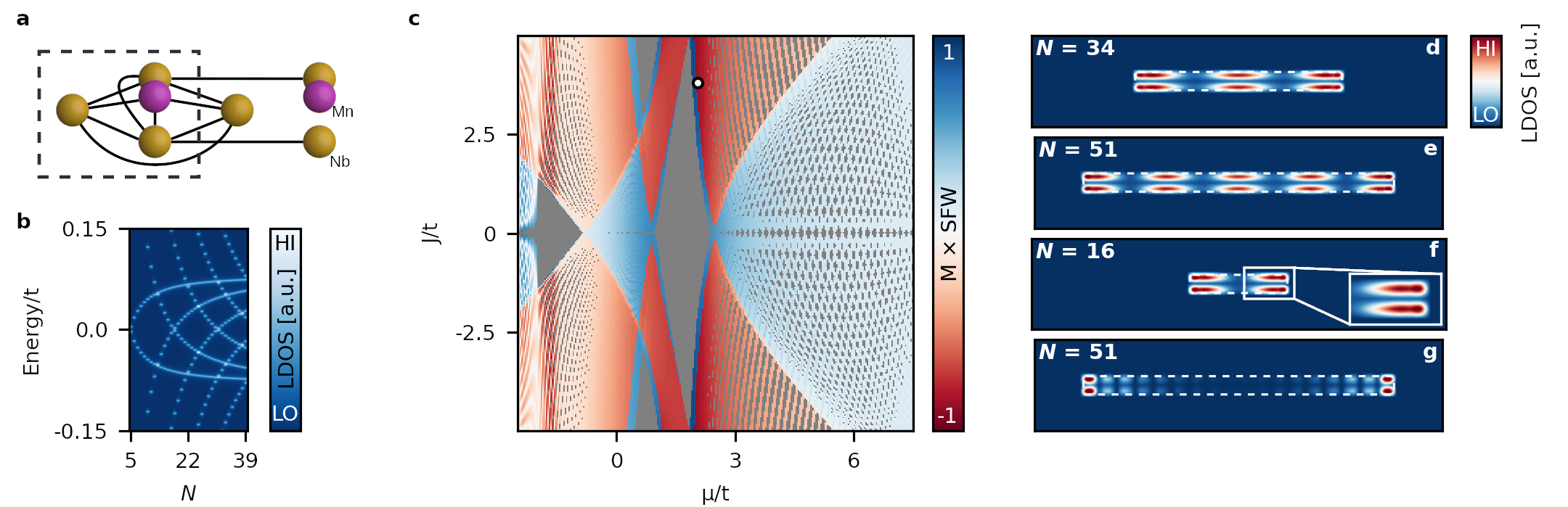} 
\caption{\textbf{Properties of simplified model for Majorana modes with side features. a,} Definition of unit cell and hoppings of the simplified model. \textbf{b,} Evolution of the side features with chain length $N$ in analogy to the experimental Fig.\,\ref{fig:fig3}\,b. Parameters: $(\Delta, \alpha, \mu, J)=(0.05, 0.01, 2.05, 3.4)t$. \textbf{c,} Representative topological phase diagram for $\Delta= 0.05t$, $\alpha=0.01t$ and $N=51$. Topologically non-trivial (trivial) regions are shown in red (blue), and are multiplied with the {\it side feature weight} (SFW) which quantifies how much spectral weight is accumulated on the side-Nb atoms (see SI section S3.B for details). Grey regions correspond to lowest energy states with $E>0.04\,t$, which are thus not of interest.
The white dot marks the parametric location of Fig.1\,e. \textbf{d,} Real-space zero-energy LDOS$(x,y)$ with side features revealing a periodic pattern reminiscent of a Mn$_{34}$ chain on Nb(110)\,\cite{schneider-21np}. Parameters: $(\Delta, \alpha, \mu, J)=(0.05, 0.01, 2.05, 3.4)t$. \textbf{e,} Real-space zero-energy LDOS$(x,y)$ with side features revealing a periodic pattern reminiscent of a Mn$_{51}$ and Mn$_{52}$ chains on Nb(110), {\it c.\,f.}\,Fig.\,\ref{fig:fig3}\,a. Parameters: $(\Delta, \alpha, \mu, J)= (0.05, 0.01, 2.05, 3.8)t$. \textbf{f,} Real-space zero-energy LDOS$(x,y)$ with side features with reduced periodic pattern in the bulk reminiscent of a Fe$_{16}$ chain on Nb(110), {\it c.\,f.}\,Fig.\,\ref{fig:fig3}\,c. Parameters: $(\Delta, \alpha, \mu, J)=(0.1, 0.05, 2, 4.9)t$. \textbf{g,} Real-space zero-energy LDOS$(x,y)$ with side feature only at the chain ends, reminiscent of the observed double-eye feature for Fe chains on Pb(110)\,\cite{feldman-17np286}. Parameters: $(\Delta, \alpha, \mu, J)=(0.15, 0.1, 1.35, 6)t$. The zero-energy LDOS states shown in d--g are in the topological phase.}
\label{fig:fig4}
\end{figure*}

Next, we study the real-space electronic structure, and in particular that of the MZMs of the DFT-based 80-band model. 
For short chains, it is very difficult to find localized zero-energy states due to the tiny gap size leading to a severe hybridization of MZMs (Fig.\,\ref{fig:fig2}b). Fig.\,\ref{fig:fig2}b reveals that there is no unique structure of MZMs, making it difficult to identify topological phases solely based on the spatial structure of the zero-energy local density of states (LDOS).
However, our computation of the topological invariant in real-space\,\cite{budich-13prb075419} shows that, as long as chains exceed $\sim10$ sites, we find perfect agreement with our momentum space analysis shown in Fig.\,\ref{fig:fig2}\,a. For $\Delta>200$\,meV we find examples of zero-energy LDOS where the hybridization leads to an oscillatory pattern and, quite astonishingly, the spectral weight of the MZMs is not located in the center of the chain but instead on the Nb atoms on the side of the magnetic chain.
These side features show a striking similarity to recently reported scanning tunneling spectroscopy (STS) measurements on a Mn$_{34}$ chain on Nb(110)\,\cite{schneider-21np} ({\it i.e.}, a chain consisting of 34 Mn atoms). We show in Fig.\,\ref{fig:fig2}\,c three examples of the zero-energy LDOS corresponding to different values of $\Delta$ and different chain lengths: a Mn$_{34}$ chain with $\Delta=320$\,meV (three maxima), a Mn$_{51}$ chain with $\Delta=300$\,meV (five maxima) and a Mn$_{52}$ chain with $\Delta=270$\,meV (six maxima). 
\sr{Their topological gap sizes for open boundary conditions are $3.10$\,meV, $1.37$\,meV and $0.88$\,meV, respectively. We note that a self-consistent treatment will typically suppress $\Delta$ near the magnetic chain, reducing the effective gap by a sizeable amount. This reduction can easily be as large as a factor of 3 or 4, which would bring the theoretical gap size to the same range as the experimental gap size of $180$\,$\mu$eV \sr{of one of the multi-orbital YSR-bands} in the vicinity of the Mn[001] chains on Nb(110)\,\cite{schneider-21np} (which is a factor of 8 smaller than the superconducting bulk gap, $\Delta_{\rm Nb}=1.51$\,meV).
}
We emphasize that localization is governed by the correlation length, and thus longer chains lead to stronger localization of MZMs at the chain ends (see SI section S3.D). We note that side features can be observed for several values of $\Delta$ per chain length; 
we further note that MZMs can also appear in the chain centre or combinations of side and centre atoms (see for instance Fig.\,S3 in SI section S2). It turns out that contributions from different orbitals are sensitive to the chosen value of $\Delta$, leading to a complex, non-trivial interplay of the five $d$-orbitals in the low-energy LDOS.
All theoretical LDOS plots shown in the main text are purely two-dimensional. In SI section S4 we extend our analysis by including tip trajectory and tunneling geometry and show that the 3D LDOS still contains sides features; all our conclusions remain, hence, unchanged.

Throughout the manuscript, we use the term MZM for the low-energy or zero-energy states of the topological superconducting phase. The latter is determined by the topological invariant for both infinite and finite chains. Clearly several of the considered examples of MZM are (strongly) hybridized, and in the literature alternative names such as `precursor of MZMs' have been introduced\,\cite{schneider2}. It is unknown whether hybridized MZMs are useful for braiding and thus for topological quantum computing, but that is not the topic of our paper. The aim of this paper is to demonstrate that a DFT-based modeling leads to a topological superconductor with low-energy states which are exceedingly similar to what has been observed in the experiments.

%
%

\begin{figure*}[t!]
\centering
\includegraphics[width=0.97\linewidth]{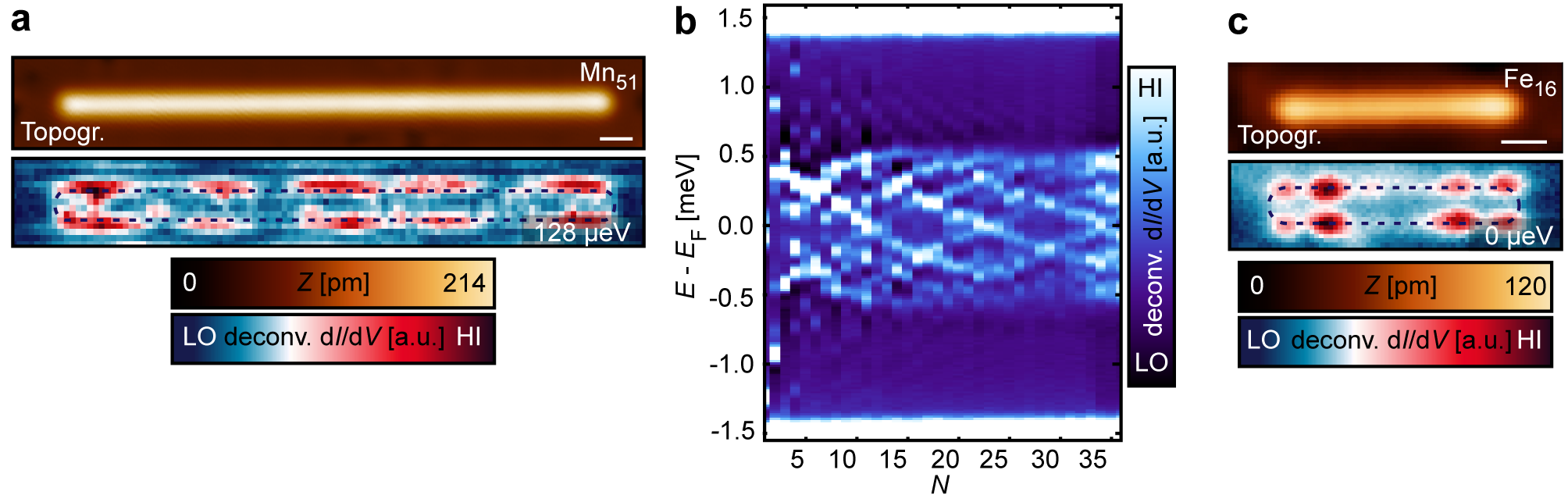} 
\caption{
\textbf{Experimental observation of low-energy side features in linear atomic Mn and Fe chains on Nb(110). a,} Constant-current image (top panel) and deconvoluted $dI/dV$ map (bottom panel) at $E = 128\,\mu$eV for a Mn$_{51}$ chain along the [001] direction. The white bar corresponds to 1\,nm. The apparent extent of the chain from the upper panel is marked by the dark dashed lines in the $dI/dV$ map. \textbf{b,} Deconvoluted $dI/dV$ spectra measured on the side of Mn$_N$ chains’ one end while additional Mn atoms were attached to the chain on the other end, thereby mainly showing the evolution of the side features with chain length $N$ ({\it c.\,f.} Refs.\,\cite{schneider-21np,schneider2}). \textbf{c,} Constant-current image (top panel) and deconvoluted $dI/dV$ map (bottom panel) at zero energy for a Fe$_{16}$ chain along the [001] direction, which is structurally identical to the Mn chain in panel a. The white bar corresponds to 1\,nm. The apparent extent of the chain from the upper panel is marked by the dark dashed lines in the $dI/dV$ maps. Parameters: $V_{\rm stab} = -6$\,mV, $I_{\rm stab} = 1$\,nA, $V_{\rm mod} = 20\,\mu$V.
}
\label{fig:fig3}
\end{figure*}

\vspace{20pt}
\noindent
\textbf{Majorana zero modes with side features}.~
In order to deepen our understanding of the side features in the zero-energy LDOS, we analyze in the following a simplified model. Inspired by relevant orbitals of the 80-band DFT-based model which lead to side features (see 
SI section S2), we construct a one-orbital model which still has three “substrate” atoms carrying superconductivity and one magnetic atom (modeled by a Zeeman term), see Fig.\,\ref{fig:fig4}\,a. 
\sr{We stress that the model still contains a non-trivial hopping structure which could not have been guessed by an {\it ad hoc} toy model; the multi-band DFT input is key to derive and justify this `simplified' model.
}
We find for large gaps localized MZMs in the center of the chain ends (either on the magnetic or on the substrate atom), in agreement with previous modeling. However, when reducing the gap size (in an attempt to simulate the small experimental gap size), we find that the spectral weight of the MZMs drops to almost zero on the magnetic atoms. Moreover, spectral weight appears on the side-Nb atoms, in agreement with the DFT-based model, reminiscent of previous experimental findings\,\cite{schneider-21np}.
A logical explanation for the formation of side features is as follows: the gap size corresponds to the strength of the superconductivity and thus to the proximity effect. In case of a large gap, {\it i.e.}, of a significant superconducting proximity effect, superconductivity can be induced into the magnetic atoms and the MZM can be localized on the magnetic atoms at the chain ends. If the gap is small, however, the proximity induced superconductivity competes with the magnetism of the Mn atoms, leading to suppressed superconducting correlations, and the MZMs are pushed away and localize along both sides instead. 

The simplified model allows us to tune various parameters such as Rashba spin-orbit coupling, magnetic moment, chemical potential and superconducting pairing amplitude. We have analyzed the corresponding phase diagram and computed the fraction of the zero-energy LDOS which is localized on the side next to the magnetic chain.
40 percent of the topological region of the phase diagram Fig.\,\ref{fig:fig4}\,c turns out to exhibit notable side features, and ten percent of the examined parameter space possess 70-99 percent of the total spectral weight on the sides of the chain. 
We have also analyzed a variant of the model with an extended substrate in both $x-$ and $y-$direction to rule out confinement effects perpendicular to the chain. We find that side features persist regardless of the size of the substrate or its boundary conditions (see SI section S3.F for details). These findings emphasize that side features in the low-energy LDOS are not the exception but can be rather common. 
The oscillating behavior of the low-energy modes is a more ordinary hybridization effect, which has previously been discussed in the context of other systems\,\cite{klinovaja-12prb085408,stanescu-13prb094518,albrecht-16n206,deng-16s1557,deng-18prb085125}. 
We find that the smallest considered gap sizes lead to an equal-distributed pattern of oscillatory maxima along the chain (Fig.\,\ref{fig:fig4}\,e, as found for the Mn$_{51}$ chain in Fig.\,\ref{fig:fig2}\,c). 
We also computed the evolution of side-feature states with chain length $N$ displaying an oscillatory behavior of the low-energy states, see Fig.\,\ref{fig:fig4}\,b.

When slightly increasing the gap size, these maxima disappear in the middle of the chain, but persist closer to the chain ends (an example with two such maxima at the chain ends is shown in Fig.\,\ref{fig:fig4}\,f). Finally, we also find situations where MZMs are confined to the ends of the chains, still showing the side feature (Fig.\,\ref{fig:fig4}\,g). We obtain a clear physical picture which emerges from our analysis: 
(i) when the substrate is included (via the three extra atoms per unit cell) and the gap is small, MZMs are forced to localize beside the magnetic chain on the substrate; 
(ii) for the smallest gap sizes considered, hybridization of MZMs are so strong that we obtain a periodic pattern of maxima in the zero-energy LDOS [see Fig.\,\ref{fig:fig4}\,d and e], as previously observed for the {\it ab initio} computed Mn$_{51}$ chain in Fig.\,\ref{fig:fig2}\,c and for a measured Mn$_{34}$ chain in Ref.\,\cite{schneider-21np}. This is in agreement with a very large coherence length $\xi \sim v_F/\Delta_{\rm gap}$ exceeding the chain length (see SI section S3.D). Increasing the gap size reduces $\xi$ and leads to a decaying envelope of the wavefunction, such that spectral weight in the bulk of the chain is suppressed. 

Larger gap sizes result in even stronger localization of the MZMs [see Fig.\,\ref{fig:fig4}\,g], showing a remarkable similarity to what was observed for an Fe chain on a Pb(110) substrate\,\cite{feldman-17np286}. We have computed coherence lengths and low-energy LDOS for small, medium and large gap sizes for extended parameter ranges, all in support of our previous conclusions (see SI sections S3.C and S3.D).
While our {\it ab initio} model does not describe the Fe/Pb(110) system, we can still speculate that the topological low-energy features might not depend on the specific system: there is clearly a strong similarity of the observed side feature with shorter coherence length (Fig.\,\ref{fig:fig4}\,g) and the “double eye” seen in experiments\,\cite{feldman-17np286}.
Due to the topological character we propose that the double eye feature found in Fe/Pb(110) has the same origin as the side features reported and modeled in this paper.

%
%

\vspace{20pt}
\noindent
\textbf{Experiment}.~
To test the theoretical findings presented above, we studied linear chains of Mn and Fe assembled via lateral atom manipulation techniques on a clean Nb(110) surface\,\cite{schneider-21np,schneider2,schneider3} (see Refs.\,\cite{method,schneider-21np,schneider2} for details on sample preparation and chain construction). Fig.\,\ref{fig:fig3}\,a shows the topography of a Mn$_{51}$ chain assembled along the [001] direction (upper panel), together with the spatial form of $dI/dV$ for a state close to zero energy. The side features visible in $dI/dV$ exhibit a close spatial similarity with those that exist in the theoretical LDOS shown in Figs.\,\ref{fig:fig2}\,c and \ref{fig:fig4}\,e (previously reported side features, such as those shown in Fig.\,3 of Ref.\,\cite{schneider-21np}, also exhibit a close similarity with Figs.\,\ref{fig:fig2}\,c and \ref{fig:fig4}\,d). The energy at which these side features are visible varies with the chain length $N$\,\cite{schneider-21np}; 
the evolution of the side features’ energies with $N$ is shown in Fig.\,\ref{fig:fig3}\,b. Notably, the lowest energy state exhibiting these side features does not converge to $E = 0$ for chain lengths up to $N = 36$ but rather oscillates around zero energy. This result resembles the theoretical predictions shown in Fig.\,\ref{fig:fig4}\,b (see also SI section S3.E) for a system with a small superconducting gap, and a chain length that is smaller than the coherence length. The combination of these two renders the experimental resolution of a topological gap exceedingly difficult. Indeed, our numerical simulations show that for the case shown in Fig.\,\ref{fig:fig4}\,b, the chain length has to be significantly larger 
before a topological gap can be resolved and the Majorana mode is clearly separated from the bulk states.

To study the universality of the observed side features, we also investigated  the low-energy properties of a Fe$_{16}$ chain along the [001] direction. This chain is structurally identical (see upper panel in Fig.~\ref{fig:fig3}\,c) to the Mn chains presented in Fig.\,\ref{fig:fig3}\,a and Ref.\,\cite{schneider-21np}, and predicted to have a ferromagnetic ground state as well\,\cite{laszloffy-21rXiv2106.09977}.
It also exhibits an enhanced zero energy $dI/dV$ at both sides of the chain ends (lower panel in Fig.\,\ref{fig:fig3}\,c). The notable similarity of these results with the theoretical calculations presented in Fig.\,\ref{fig:fig4}\,f, suggests a common origin of the side features observed for the  Mn$_{51}$ and Fe$_{16}$ chains on Nb(110).

%
%
\section{Discussion}


The side features of the Mn chains were previously ascribed to bands formed by hybridization of Yu-Shiba-Rusinov (YSR) states of particular orbital character of the individual Mn impurities\,\cite{schneider-21np,beck-21nc2040}. Individual Fe atoms do not exhibit clear signatures of multi-orbital YSR states\,\cite{odobesko-20prb174504,kuester-21nc1108} (see 
SI section S6 for details), due to their close proximity to the coherence peaks of the substrate.
Nevertheless, the linear Fe chains along the [001] direction, mentioned above, show an enhanced zero-energy LDOS at the sides of both ends (Fig.\,\ref{fig:fig3}\,c). This surprising observation, which cannot be explained by a linear combination of YSR states of the individual atoms (see SI section S6), shows obvious similarities with the theoretical calculations presented in Fig.\,\ref{fig:fig2}\,c and Fig.\,\ref{fig:fig4}. The interpretation of the observed side features as hybridized Majorana modes, as advocated in the theoretical part of this paper, is thus substantiated by the attempt to simultaneously explain the low-energy features for Mn and Fe chains on Nb(110).

Motivated by our analysis relating these findings with the previously discovered double-eye feature\,\cite{feldman-17np286}, we thus conclude that MZMs with side features might be a common phenomenon in MSH materials with small superconducting gap size. We note, however, counter examples do exist such as Fe/Re(0001)\,\cite{kim_toward_2018,schneider-20nc4707} and  Mn chains along [1$\bar{1}$0] on Nb(110)\,\cite{schneider2}.
%
This is in fact consistent with our analysis of the superconducting 80-band model: side features appear only for certain values of $\Delta$, and changing $\Delta$ leads to a redistribution of the relative orbital weight. Since some of the Mn $d$-orbitals support side features and others do not (see discussion of Fig.\,S3 in SI section S2), these `counter examples' are compatible with the main results of our work.
We emphasize that our results have far-reaching consequences: the presence of point-like zero-energy modes at chain ends is considered as a necessary condition for the identification of one-dimensional topological superconductors. Our work suggests that MZMs could in fact come in a large variety, ranging from delocalized, hybridized MZMs along the chain to localized MZMs on the chain.
{\color{black}What determines whether a system is topological or not is not the spatial form of the lowest energy state, but the calculated value of the topological invariant.
}
Our findings raise the important question whether MZMs with side features can be used for braiding and topological quantum computing, emphasizing the necessity to perform further theoretical and experimental work on one-dimensional MSH structures.

%
%

\section*{Methods}

\noindent
\textbf{Density Functional Theory Calculations}~
We are using fully relativistic all electron DFT calculations based on the full potential local orbital (FPLO) basis\,\cite{koepernik-99prb1743}    
in combination with a generalized gradient approximation (GGA) to the exchange correlation functional. We obtain low energy tight binding models using projective Wannier functions\,\cite{Eschrig2009}.   
The considered slab supercell is shown in Fig.\,1\,a.
We have fully relaxed the top half of the system while keeping the two bottom layers fixed at the Nb bulk lattice parameters and internal coordinates in order to simulate the experimental Nb(110) surface. We find that the top Nb layers contract slightly, lowering the nearest neighbour bond lengths. High symmetry positions for the placement of Mn are on top of Nb atoms and in the central hollow of the diamond arrangement of four Nb surface atoms. Energetically, the hollow position is 1.4\,eV/Mn more favorable, in agreement with experimental observations and the calculations in Ref.\,\cite{beck-21nc2040}.  
The Nb atoms forming bonds with Mn are measurably distorted out of their equilibrium Nb surface positions. 

\vspace{20pt}
\noindent
\textbf{Superconducting 80-Band Modeling}~
The projected 40-band tight-binding model is given by 
\begin{equation}
H=\sum_{i,j;o,o';\sigma,\sigma'} t^{i,o;j,o'}_{\sigma \sigma'} c_{i,o,\sigma}^\dag \, c_{j,o',\sigma'}^\pd + {\rm H.c.}
\end{equation}
including essentially all possible hopping combinations up to fifth neighbors. Hopping amplitudes $t^{i,o;j,o'}_{\sigma \sigma'}$ are given as an input from DFT, and $c_{i,o,\sigma}^\dag$ creates an electron with spin $\sigma$ at orbital $o$ at site $i$.
We add the superconducting onsite pairing term 
$H_\Delta=\sum_{i,o}\Delta c_{i,o,\up}^\dag c_{i,o,\dw}^\dag + {\rm H.c.}$ and convert the model into an 80-band Bogoliubov--de Gennes model. As stressed before, the orbital index $o$ only runs over the orbitals of the Nb atoms.
We have calculated the topological invariant\,\cite{kitaev01pu131}   
both in momentum and real space\,\cite{budich-13prb075419}    
for this model. By virtue of Fourier transformation we obtained the real-space representation, leading to matrix dimensions of $80 N$ for chain length $N$ which can be diagonalized numerically. Open boundary conditions are enforced by setting all hopping amplitudes connecting the two chain ends to zero; to make the chain mirror symmetric, the central Nb atom is removed on the left side of the chain. In the SI section S3.F, we investigate finite chains embedded in a larger substrate in order to rule out edge and confinement effects.

\vspace{20pt}
\noindent
\textbf{Scanning Tunneling Microscopy and Spectroscopy}~
The STM experiments were performed in a home-built UHV facility operated at a base temperature of $T = 320$\,mK\,\cite{wiebe-04rci4871}.   
The Nb(110) single crystal substrate was cleaned by high temperature flashes to $T > 2700$\,K, following Ref.\,\onlinecite{odobesko-19prb115437}.  
Single Mn and Fe atoms were deposited onto the cold sample surface held at $T < 7$\,K. A superconducting tip was prepared by indenting an electrochemically etched W tip into the substrate, thereby transferring large clusters of superconducting Nb onto the tip apex. STM topography images were measured in a constant-current mode with the current setpoint $I$ while applying a constant bias voltage $V_{\rm DC }$ across the tunneling junction. Differential tunneling conductance $(dI/dV)$ spectra were measured by stabilizing the tip at bias voltage $V_{\rm stab}$ and current $I_{\rm stab}$. In a next step, the feedback loop was opened and $V_{\rm DC }$ was swept between $‑4$\,mV and $+4$\,mV. The $dI/dV$ signal was measured using a standard lock-in technique with a small modulation voltage $V_{\rm mod}$ (RMS) of modulation frequency $f = 4.142$\,kHz added to $V_{\rm DC}$. The $dI/dV$ maps were acquired by recording a set of $dI/dV$ spectra on a rectangular grid. Superconducting Nb tips have been used in order to achieve a higher effective energy resolution. The measured differential tunneling conductance $dI/dV$ is thus approximately proportional to a convolution of the LDOS of the sample and the superconducting tip density of states (DOS). We show numerically deconvoluted STS data throughout the manuscript, resembling the sample’s LDOS (see Refs.\,\onlinecite{schneider-21np,schneider2} 
for details on the deconvolution process and determination of the tip gap). The Mn and Fe chains were assembled using lateral atom manipulation\,\cite{eigler-90n524,kim_toward_2018,schneider3}  
techniques at low tunneling resistances of $R \approx 12 - 60$\,k$\Omega$.

\vspace{5pt}

\paragraph*{Data availability.}
All data and materials needed to evaluate the conclusions in the paper are present in the paper and/or the Supplementary Information and are available for download at https://doi.org/10.5281/zenodo.6522954.

\vspace{5pt}

\paragraph*{Acknowledgments.}
The authors acknowledge discussions and previous collaborations with L.\ Schneider.
J.W.\ and R.W.\ gratefully acknowledge funding by the Cluster of Excellence `Advanced Imaging of Matter' (EXC 2056 - project ID 390715994) of the Deutsche Forschungsgemeinschaft (DFG). P.B., J.W.\ and R.W.\ acknowledge support by the DFG via SFB 925 – project ID 170620586. R.W.\ gratefully acknowledges financial support from the European Union via the ERC Advanced Grant ADMIRE (project No. 786020). M.S.\ acknowledges support by Grant-in-Aid for JSPS Fellows No.\ JP21J12095. S.R.\ acknowledges support from the Australian Research Council through Grant No.\ DP200101118.

\vspace{5pt}

\paragraph*{Author contributions.}
M.S. and H.O.J. performed the DFT calculations and derived the 40-band tight-binding model. D.C., E.M., D.K.M. and S.R. performed all other theoretical modeling. P.B., R.W. and J.W. conceived the experiments. P.B. performed the measurements and analysed the experimental data together with J.W. S.R. wrote the manuscript, with contributions from all authors. All authors contributed to the discussions and to correcting the manuscript.

\vspace{5pt}

\paragraph*{Competing Interests.}
The authors declare no competing interests.

\bibliography{MnNb-sidefeatures}

\end{document}


\title{Supplementary Information:\\
Majorana modes with side features in magnet-superconductor hybrid systems}

\author{Daniel Crawford}
\affiliation{School of Physics, University of Melbourne, Parkville, VIC 3010, Australia}

\author{Eric Mascot}
\affiliation{Department of Physics, University of Illinois at Chicago, Chicago, IL 60607, USA}

\author{Makoto Shimizu}
\affiliation{Department of Physics, Okayama University, Okayama 700-8530, Japan}

\author{Philip Beck}
\affiliation{Department of Physics, University of Hamburg, D-20355 Hamburg, Germany}

\author{Jens Wiebe}
\affiliation{Department of Physics, University of Hamburg, D-20355 Hamburg, Germany}

\author{Roland Wiesendanger}
\affiliation{Department of Physics, University of Hamburg, D-20355 Hamburg, Germany}

\author{Harald O.\ Jeschke}
\affiliation{Research Institute for Interdisciplinary Science, Okayama University, Okayama 700-8530, Japan}

\author{Dirk K.\ Morr}
\affiliation{Department of Physics, University of Illinois at Chicago, Chicago, IL 60607, USA}

\author{Stephan Rachel}
\affiliation{School of Physics, University of Melbourne, Parkville, VIC 3010, Australia}



\maketitle

\section*{Supplementary Note 1. Density Functional Theory calculations}\label{dft} 

In the main text, the orbital-resolved electronic density of states (DOS) of the $d$-orbitals of the Mn atoms is shown in Fig.\,1\,c and e. Here we also show the total electronic DOS in Supplementary Fig.\,\ref{fig:DFT-DOS}, separated into spin-up (blue) and spin-down (red) components. The previously presented Mn $d$-orbitals are shown as black thick lines (increased by a factor 5 for clarity). The black thin horizontal line marks the Fermi energy.

\begin{figure}[h!]
\centering
\includegraphics[width=0.8\columnwidth]{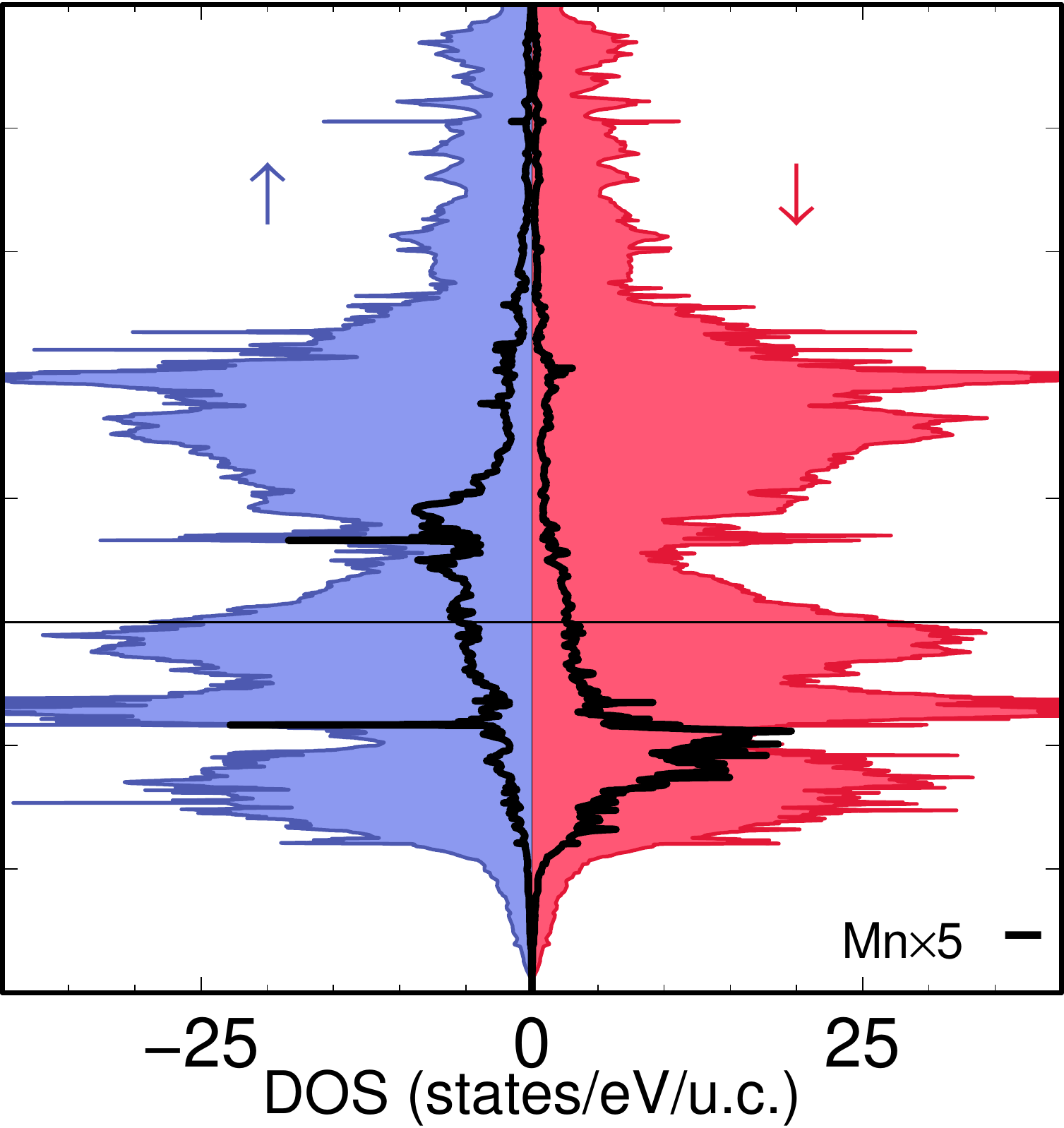}
\caption{
\textbf{Electronic DOS for the supercell with $\boldsymbol{d_{\rm Mn-Mn}=3.3}${\AA}
chains on Nb(110).} The calculation is performed with generalized gradient approximation and spin-orbit coupling (GGA+SO) with
ferromagnetically aligned magnetic moments of Mn and quantization axis $c$,
representing the DFT ground state of the system. Blue and red DOS stand
for spin minority (up arrow) and majority (down arrow), respectively. The Mn DOS
is shown by black lines and multiplied by 5 for clarity.}
\label{fig:DFT-DOS}
\end{figure}

%
%
\section*{Supplementary Note 2. 40-band tight binding and Superconducting 80-band models}\label{band-model} 

\begin{figure}[b]
\centering
\includegraphics{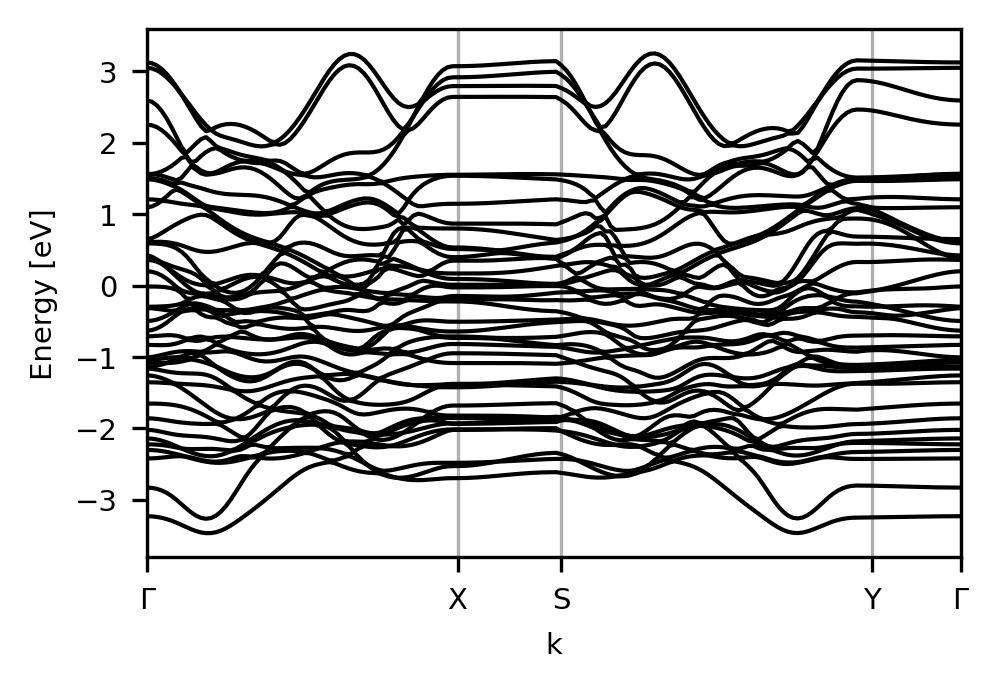}
\caption{\textbf{Normal state band structure of 40-band model.} We observe flat bands at low energies when varying 
\(k_y\), \ie between $X$ and $S$ and between $Y$ and $\Gamma$. A purely 1D model with \(\vec{k} = (k_x, 0)\) is thus justified.}
\label{fig:40band-spectra}
\end{figure}

\begin{figure}[t!]
\centering
\includegraphics[width=\columnwidth]{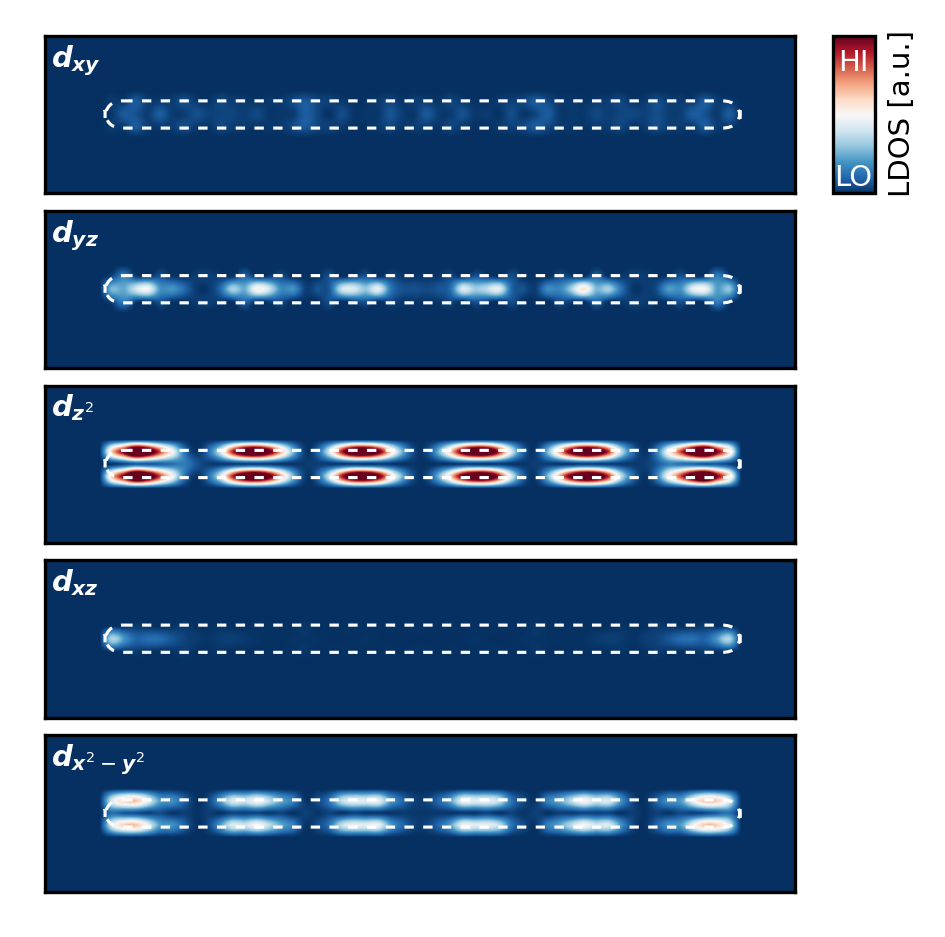}
\caption{\textbf{Orbital-resolved real-space LDOS plots at zero energy.} The data shown here is the same as in Fig.\,2\,c of the main text (for Mn$_{52}$, $\Delta=270$\,meV), but resolved by the contributions of the five Mn $d$ orbitals. 
Theoretical LDOS is always evaluated in the plane of the chain.}
\label{fig:40band-orbitals}
\end{figure}

\begin{figure*}
\centering
\includegraphics{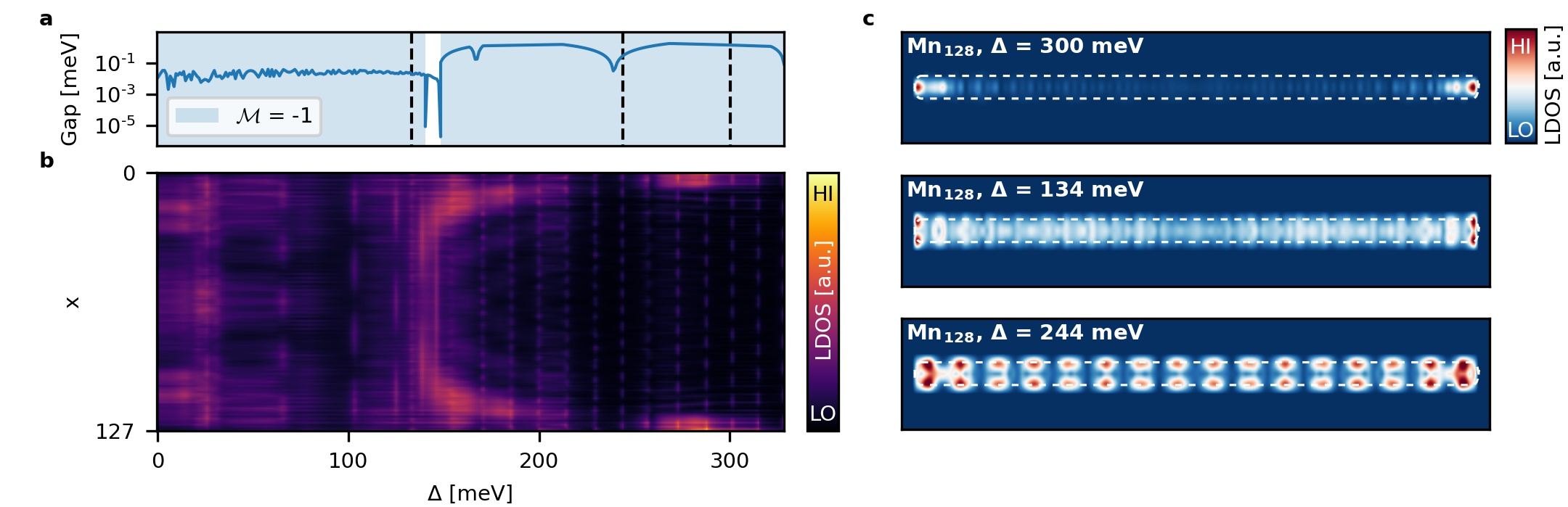}
\caption{\textbf{Analysis of DFT-based superconducting 80-band model for $N=128$.} \textbf{a,} Topological phase diagram as a function of the bare superconducting amplitude $\Delta$ (topological region in blue with $\mathcal{M}=-1$, and white is trivial). The gap size is shown as the blue curve. \textbf{b,} LDOS as a function of lattice site along $x$ and $\Delta$ (the $y$-dependence has been summed up such that LDOS appears one-dimensional). \textbf{c,} Examples of zero-energy LDOS(x,y) for various values of $\Delta$, as indicated by the dashed vertical lines in a, with pronounced side features. Chain length $N=128$, the value of $\Delta$ is given within the panels. 
}
\label{fig:40band-long-chain}
\end{figure*}

The projected 40-band tight-binding model describes the $d$-orbitals of four atoms per unit cell; one atom is on top (the ``Mn atom'', shown in purple in Fig.\,1a of the main text) and three underneath, one in the center and one on both sides (the ``Nb atoms'', shown in yellow in Fig.\,1a of the main text). The model contains complex hoppings up to fifth neighbors being spin-imbalanced due to the spin polarization.
The 40-band model contains extra hoppings in the \(y\) direction stemming from the original DFT-based model. We remove
these hoppings to work in a purely-1D system. We justify this by
inspecting the 2D spectra (Supplementary Fig.\,\ref{fig:40band-spectra}). The flat
bands in the \(k_y\) direction indicate the absence of any dispersion, thus allowing us to treat the model as one-dimensional and to dismiss any hoppings in the \(y\)-direction.

\begin{figure}[b!]
\centering
\includegraphics{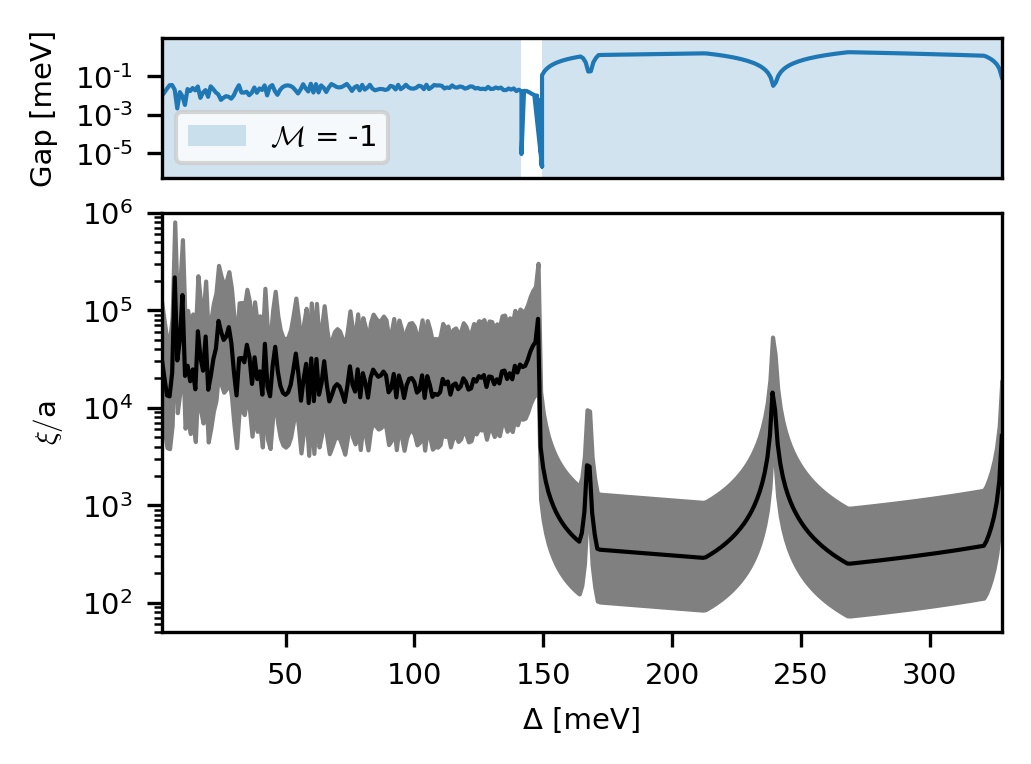}
\caption{\textbf{Spectral gap, topological invariant and coherence length of the superconducting 80-band model.}
(Top) Spectral gap and topological invariant, identical to Fig.\,2\,a of the main text. (Bottom) Coherence length $\xi=v_F / \Delta_{\rm gap}$ for  the superconducting 80-band model; $a$ is the lattice spacing. 
Due to the multi-orbital character of the model, we observed several band crossings at the Fermi energy.
The black line corresponds to the average coherence length;
the range between minimum and maximum coherence lengths are indicated by the shaded region. For
small values of \(\Delta\) much longer chains are required to avoid strong hybridization effects than
for large values of \(\Delta\). \label{fig:40band_coherence_length}}
\end{figure}

In the following, we focus on the superconducting 80-band model (see Methods in the main text). As mentioned in the main text, the peculiar side features appear only for certain values of $\Delta$, since different orbital contributions are sensitive to the choice of $\Delta$. In Supplementary Fig.\,\ref{fig:40band-orbitals} we show the orbital-resolved version of the zero-energy LDOS (always evaluated in the plane of the chain) presented in Fig.\,2\,c of the main text (Mn$_{52}$, $\Delta=270$\,meV). 
The manganese $d_{z^2}$ orbital dominates the LDOS contribution, but also the $d_{x^2-y^2}$ orbital has the same characteristic structure, albeit with weaker intensity. In contrast, the $d_{xz}$ orbital supports localized states at the chain ends only, and the $d_{yz}$ orbital reveals a periodic intensity, but in the center of the chain and not on the sides. Low-energy LDOS plots for other values of $\Delta$ appear to have the same structure as one of the other orbitals or a combination of multiple orbitals. That is, MZMs can be localized in the center of the chain, can be oscillatory in the center, or appear as side features. The LDOS plots with side features shown in Fig.\,2\,c of the main text are only one particular type of MZMs present in the topologically non-trivial range of $\Delta$.

In Fig.\,2 of the main text we displayed chain lengths matching the experimental systems. Numerically, we can explore arbitrary chain lengths (up to
computational limits). Some examples are shown for chains with \(N=128\)
sites in Supplementary Fig.\,\ref{fig:40band-long-chain}; the figure is analogous to Fig.\,2 from the main text. We find cases where the MZMs
are strongly localized at the ends of the chain; cases where the MZMs are
localized at the ends but pushed to the side of the chain; and cases where
 side features oscillate.

We compute all coherence lengths \(v_F/\Delta_{\rm gap}\) (stemming from  multiple bands) and compute the average coherence length. In Supplementary Fig.\,\ref{fig:40band_coherence_length} we show  average, smallest and largest coherence lengths as a function of $\Delta$, extracted from the band structure of the 40-band tight-binding model. In the following, we concentrate on the average coherence length only. In general, coherence lengths
are long and we expect strong hybridization effects of MZMs. Only for relatively long chains
(\(> 100\) sites) and for large \(\Delta\) (\(> 150\) meV) do well-localized
end states appear.

%
%
\section*{Supplementary Note 3. Simplified single-orbital model}\label{toy-model} 

We give here the formal definition of the simplified model introduced in the
main text, in technical detail. We then present some extra results:
 phase diagrams,  spectra, and  a discussion on
coherence lengths in this model. Finally, we present a modified version of the model where we have extended the substrate around the magnetic chain.

\subsection{Definition}\label{definition} 

We start with the standard Shiba chain model\,\citep{li-14prb235433}, 
and extend to a four atom unit cell. The
operators \(c_x^{(i)} = (c_{x,\uparrow}^{(i)}, c_{x,\downarrow}^{(i)})\)
annihilate an atom of type \(i=0,1,2,3\) on site \(x\); \(i=0\) refers
to the magnetic adatom, \(i=1\) to the substrate Nb atom in the centre,
and \(i=2,3\) to the Nb substrate atoms on either side of the adatoms. The
Hamiltonian is
\begin{equation}\label{toy}
H = H_t + H_\mu + H_\alpha + H_\Delta + H_J.
\end{equation}
\(H_t\) is the hopping Hamiltonian. In the general case, hopping
amplitudes 
$t_{ij}$ are defined in Supplementary Fig.\,\ref{fig:hopping_diagram}.
\begin{equation}\label{ham-t}
H_t = \sum_{i,j,x} ( t_{ij} c_x^{(i)\dagger} c_x^{(j)} 
+ t_{ij} c_x^{(i)\dagger} c_{x+1}^{(j)} + \text{H.c.} ).
\end{equation}
with \(i,j \in [0,1,2,3]\). The first term describes hopping within the unit cell (thus \(i \neq j\)), and the second term between neighboring unit cells.
\(H_\mu\) is the chemical potential,
\begin{equation}
H_\mu = \sum_{i,x}- \mu c^{(i)\dagger}_x c^{(i)}_x\ .
\end{equation}

\begin{figure}[b!]
\centering
\includegraphics[width=0.95\columnwidth]{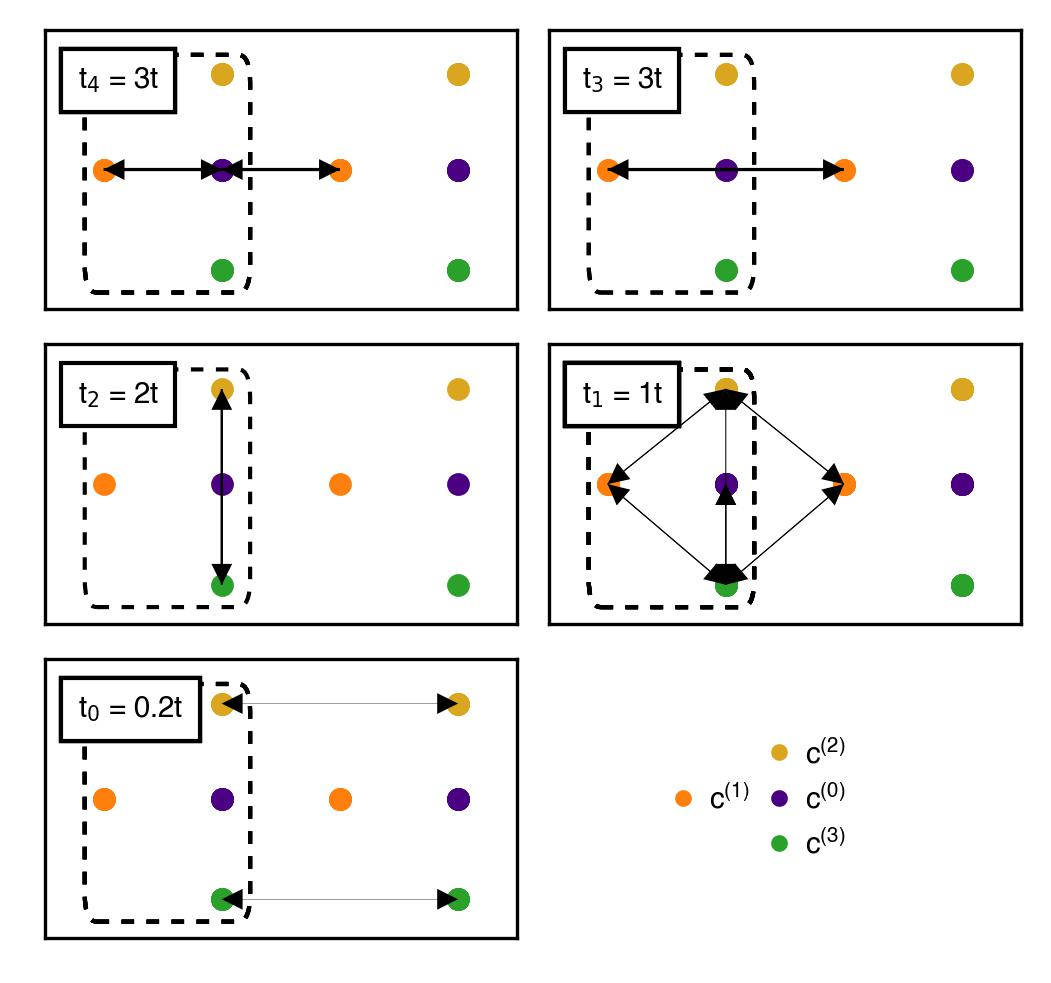} 
\caption{\textbf{Illustration of all hopping processes in the single-orbital model Eq.\,\eqref{toy}.} Arrow thickness indicates the real hopping amplitudes between the indicated atoms, defining Eq.\,\eqref{ham-t}. The dashed
rectangle indicates the unit cell. 
\label{fig:hopping_diagram}}
\end{figure}

\(H_\alpha\) is Rashba spin-orbit coupling. In the general case, Rashba
amplitudes are defined by a matrix
\(\alpha_{ij} (x, x+1) = \alpha_{ij}\). All hoppings are associated with
\(i \alpha \sigma_y\) with $\alpha$ being the Rashba amplitude. 

\begin{align}
H_\alpha = \sum_{i,j,x} (i \alpha_{ij} c_x^{(i)\dagger} \sigma_2 c_x^{(j)} + i \alpha_{ij} c_x^{(i)\dagger} \sigma_2 c_{x+1}^{(j)} + H.c.)\ .
\end{align}

\(H_\Delta\) represents \(s\)-wave onsite superconductivity, and is restricted to the
substrate sites \(i\in [1,2,3]\),
\begin{equation}
H_\Delta = \Delta \sum_{i,x} c_{x,\uparrow}^{(i)\dagger} c_{x,\downarrow}^{(i)\dagger} + \text{H.c.}
\end{equation}
\(H_J\) is the ferromagnetically aligned Zeeman field, and is restricted
to the adatoms $i=0$,
\begin{equation}
H_J = J \sum_x c^{(0)\dagger}_x \sigma_3 c^{(0)}_x.
\end{equation}

By choosing different combinations of hoppings (both spin-preserving, \ie $t_{ij}$, and spin-flip hoppings, \ie $\alpha_{ij}$) we have generated several variants of Eq.\,\eqref{toy}; all these models
generically show side features.

In the following, we focus on a particular version of Eq.\,\eqref{toy} with the hoppings defined in Supplementary Fig.\,\ref{fig:hopping_diagram}; Rashba spin-orbit coupling lies exclusively along the same arrows. 

\subsubsection{1D-Rashba  vs.\ 2D-Rashba SOC}\label{rashba-1d-vs-2d} 

Because the unit cell of our 1D model is 2D, \ie has spatial dependence in both $x$ and $y$ directions, the definition of Rashba spin-orbit coupling is ambiguous. We could choose \(i\alpha \sigma_x\) in \(y\) direction,
\(i \alpha \sigma_y\) in \(x\) direction, and a combination for diagonal
hoppings (2D choice). Alternatively, we could uniformly choose \(i \alpha \sigma_y\)
in all directions, and fix the sign convention of ``\(+\)'' for \(+y\) and ``\(-\)''
for \(-y\) (1D choice).

\begin{figure}[t]
\centering
\includegraphics[width=\columnwidth]{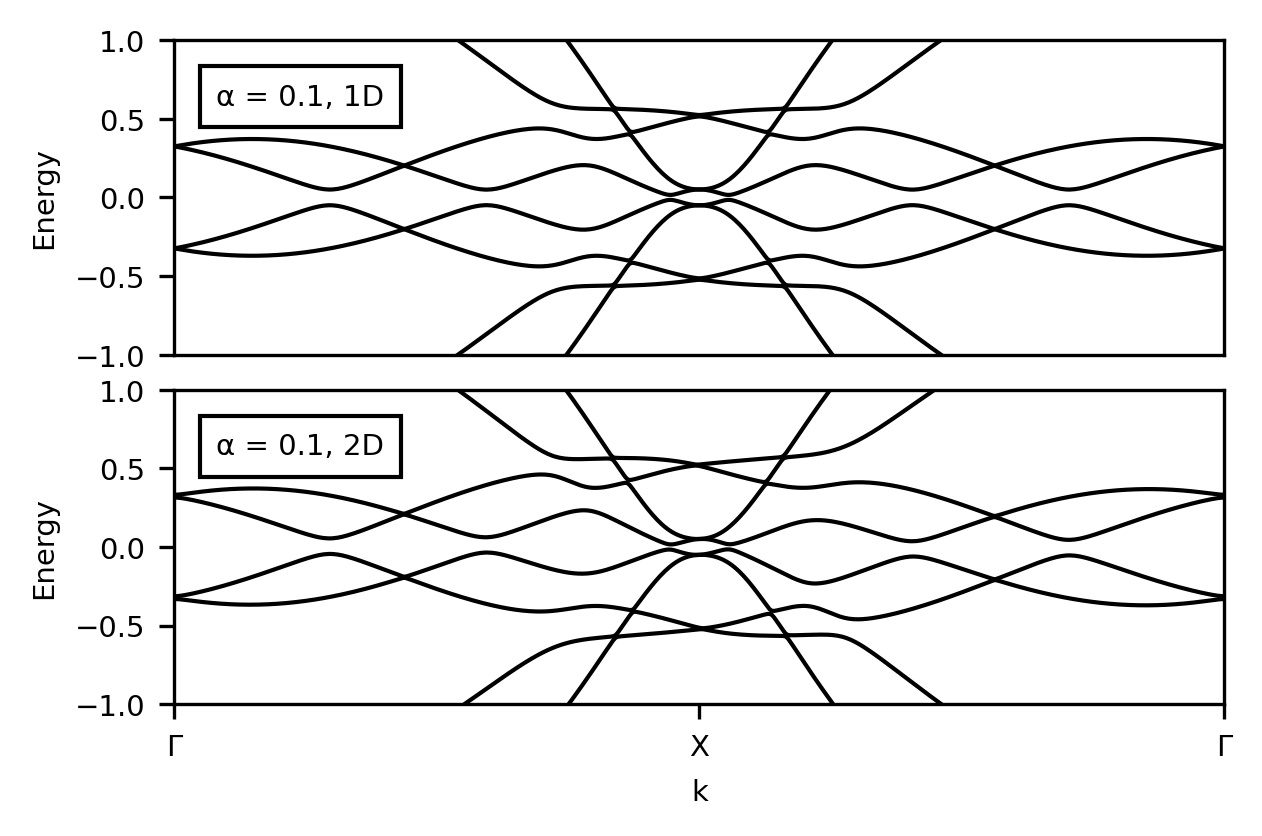} 
\caption{\textbf{Comparison of Rashba Hamiltonian convention for large Rashba SOC}.
Superconducting spectra for 1D-Rashba choice (top) and 2D-Rashba choice (bottom) for \(\alpha=0.1 t\).
\label{fig:rashba_spectra_large}}
\end{figure}
\begin{figure}[t]
\centering
\includegraphics[width=\columnwidth]{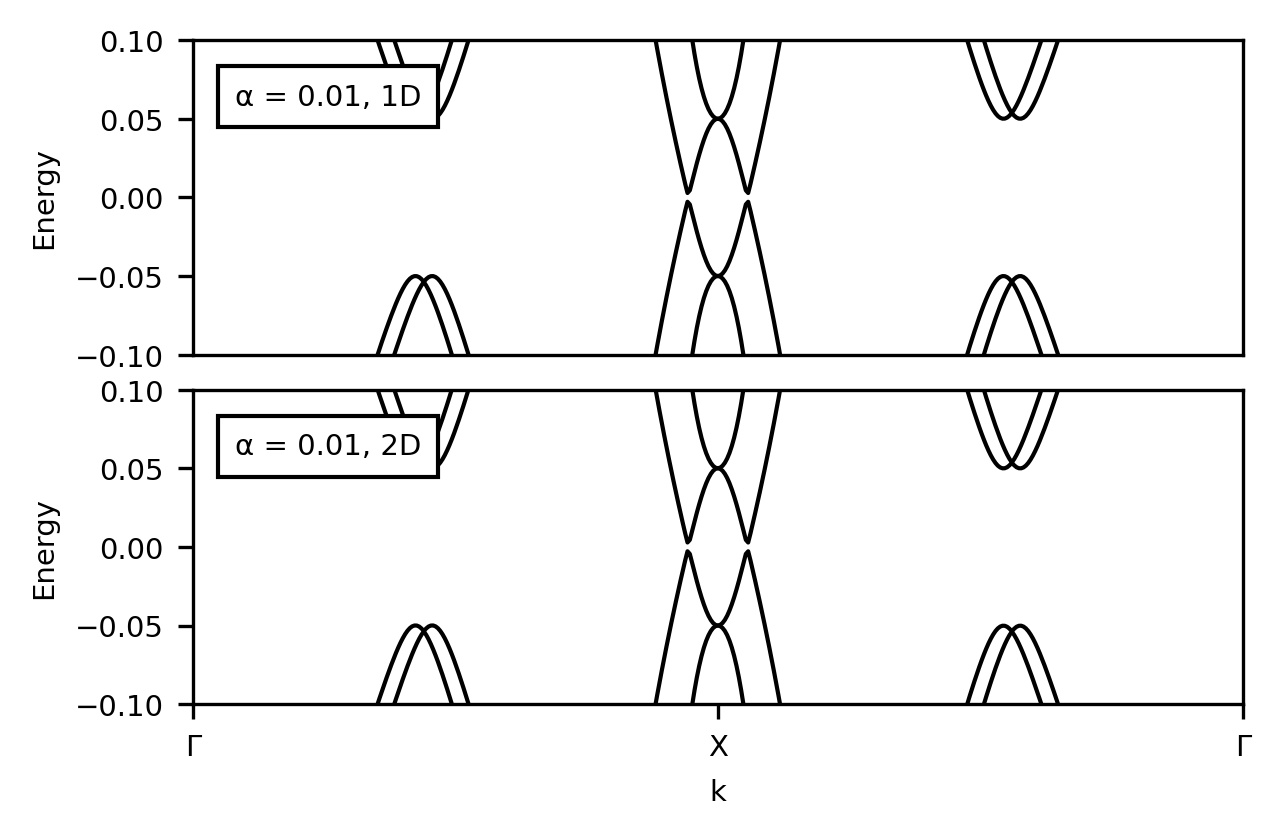} 
\caption{\textbf{Comparison of Rashba Hamiltonian convention for small Rashba SOC.} 
Superconducting spectra for 1D-Rashba choice (top) and 2D-Rashba choice (bottom) for \(\alpha=0.01 t\).
\label{fig:rashba_spectra_small}}
\end{figure}

\begin{figure*}
\centering
\includegraphics[width=0.98\linewidth]{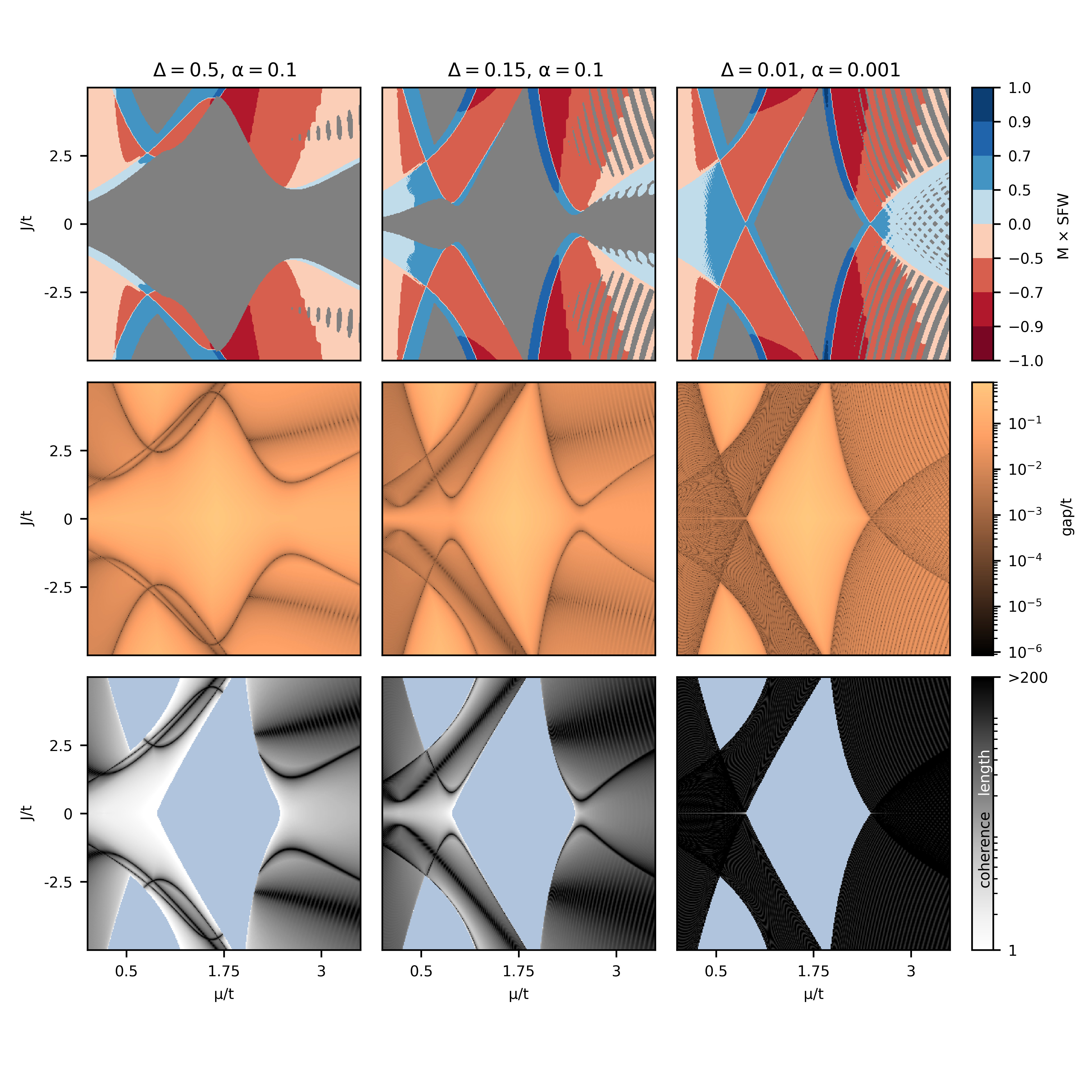} 
\caption{\textbf{Topological phase diagram, side-feature weight, gap size and coherence length.}
All panels correspond to model\,\eqref{toy} with $\Delta$ and $\alpha$ as specified above the top panels for the entire column, and are shown as a function of $J/t$ and $\mu/t$. The first column corresponds to the largest, and the third to the smallest gap sizes.
\textbf{First row:} Topological invariant $M$ multiplied by side feature weight $S(\ket{\psi})$, where $\ket{\psi}$ is the state with energy closest to $E=0$. Red (blue) regions are topologically nontrivial (trivial). Strong intensities indicate large side feature weights. Note that a small gap is required to reach high side feature weights. \sr{Grey regions correspond to lowest energy states with $E>0.04\,t$, which are thus not of interest.}
\textbf{Second row:} Gap diagrams. Note that the gap closing lines can correspond to phase transitions in the first row. Small gap sizes ($10^{-2} t \ldots 10^{-5} t$) are associated with large side feature weight.
\textbf{Third row:} Coherence length $\xi=v_F/\Delta_\text{gap}$ of the corresponding low-energy state. Light blue regions have insulating normal state, thus the coherence length is ill-defined. Note that $\xi$ can be significantly larger than 200 sites. 
System size for all panels is $N=51$.}
\label{fig:phasediagram}
\end{figure*}

For the 2D choice the Rashba part of the Hamiltonian is given by
\begin{align}
\nonumber
H_\alpha = &\sum_{i,j \,\in \{0,1\},x}
\!\!\!\!(i \alpha_{ij} c_{x}^{(i)\dagger} \sigma_2 c_x^{(j)} + i\alpha_{ij} c_{x}^{(i)\dagger} \sigma_2 c_{x+1}^{(j)} ) \\
\nonumber
+&\sum_{i=1,j\,\in\{2,3\},x}
\!\!\!\!(i \alpha_{ij} c_{x}^{(i)\dagger} \sigma_1 c_x^{(j)} + i\alpha_{ij} c_{x}^{(i)\dagger} \sigma_2 c_x^{(j)} ) \\
\nonumber
+&\sum_{i\,\in\{0,1\},j\,\in\{2,3\},x}
\!\!\!\!(i \alpha_{ij} c_{x}^{(i)\dagger} \sigma_1 c_{x}^{(j)} + i\alpha_{ij} c_{x}^{(i)\dagger} \sigma_2 c_{x+1}^{(j)})\\
+&\sum_{i,j\,\in\{0,2,3\},x}
\!\!\!\!(i \alpha_{ij} c_{x}^{(i)\dagger} \sigma_1 c_x^{(j)} ) + \text{H.c.}
\end{align}

Hoppings in the \(x\) direction are associated with \(\sigma_2\), in the
\(y\) direction with \(\sigma_1\), and along diagonals \(\pm x \pm y\)
with \(\pm i \sigma_1 \pm i \sigma_2\).

We compare superconducting
spectra under the two different implementations of Rashba SOC (Supplementary Fig.\,\ref{fig:rashba_spectra_large} and Supplementary Fig.\,\ref{fig:rashba_spectra_small}) and find the spectra are almost
identical. Hence, the choice does not matter. Also the eigenstates are 
 similar. In order to keep the model as simple as possible, we chose the 1D-Rashba convention in the main text.


\subsection{Methods}\label{methods} 

To quantify the presence of side features, we define a simple quantity,
the \emph{side feature weight} (SFW), specifically for states closest to $E=0$. Consider the $E=0$ LDOS,

\begin{equation}
\rho^{(i)}_x = {\rm LDOS} (c_x^{(i)})\ .
\end{equation}
We then define the side feature weight of the state
\(\ket{\psi}\),

\begin{equation}
S(\ket{\psi}) = \frac{\sum_x^L \rho^{(2)}_x}{\sum_x^L \rho^{(0)}_x + \sum_x^L \rho^{(1)}_x + \sum_x^L \rho^{(2)}_x}.
\end{equation}

Here we sum over the spatial position \(x\) from 0 to chain length
\(L-1\) (unless otherwise specified \(L=51\) sites). Typically we present
side feature weight as a function of the free parameters \(\mu\) and
\(J\), with fixed \(\Delta\) and \(\alpha\). For easy reading we
multiply the side feature weight by the topological index, $+1$ ($-1$) corresponding to the topologically trivial (non-trivial) phase. Negative side feature weights correspond to the topologically superconducting regime, \ie the zero-energy state represents a MZM with side features. The side feature weight is shown in the
phase diagrams of the following subsection.

\subsection{Phase diagrams}\label{phase-diagrams} 

\begin{figure}[t!]
\centering
\includegraphics{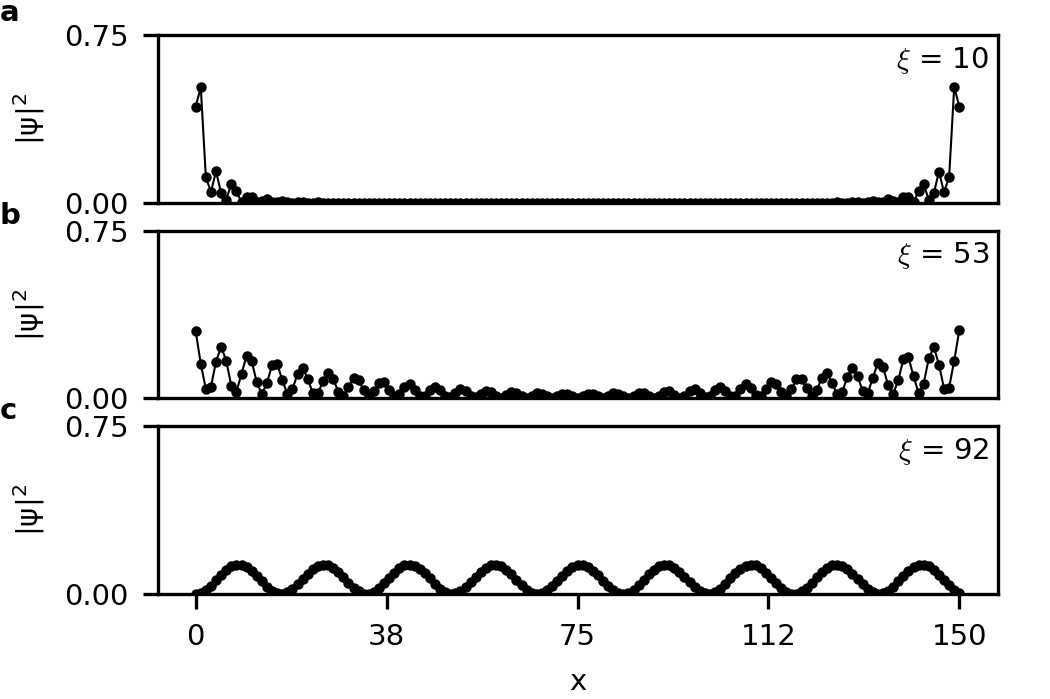} 
\caption{\textbf{Examples of wavefunctions $\bs{|\psi|^2}$ with different coherence lengths.} Shown is $|\psi|^2$ of model \eqref{toy}, summed over atoms and spin, lowest-energy state for OBC in the topological regime, thus corresponding to (hybridized) MZMs. The coherence length $\xi$ is shown in each figure panel.  \textbf{a,} Large gap regime; parameters used $(\Delta, \alpha, \mu, J)=(0.15, 0.1, 1.35, 6)t$. \textbf{b,}  Medium gap regime; parameters used $(\Delta, \alpha, \mu, J)=(0.15, 0.1, 2.4, 4)t$. \textbf{c,} Small gap regime; parameters used $(\Delta, \alpha, \mu, J)=(0.01, 0.05, -2, 4)t$. Smaller gaps are associated with longer coherence lengths and so the
MZMs hybridize and oscillate in the chain bulk. \label{fig:coh_len_wf}} 
\end{figure}

We present a representative phase diagram in Fig.\,3\,c of the main text. It turns out
that the specific parameters chosen are not too important: as long as the gap is
small we can find side features. We show phase diagrams and gap diagrams
for several parameters in the first and second row of Supplementary Fig.\,\ref{fig:phasediagram}, where different columns correspond to large, medium and small gap sizes. 
For ``large gaps'' (left column) 65\% of the entire phase diagram shows notable side features (SFW $> 0.5$), and about 9\% of the topological region displays significant side features (SFW $> 0.7$).
For ``medium gaps'' (middle column) 75\% of the entire phase diagram shows notable side features (SFW $> 0.5$), and about 11\% of the topological region displays significant side features (SFW $> 0.7$).
For ``small gaps'' (right column) about 75\% of the entire phase diagram shows notable side features (SFW $> 0.5$), and about 15\% of the topological region displays significant side features (SFW $> 0.7$).
Of course the topological phase is the same for all chain
lengths and it is, hence, sufficient to show one case ($N=51$). We note that the side-feature weight is clearly chain-length dependent.

\begin{figure}[b!]
\centering
\includegraphics{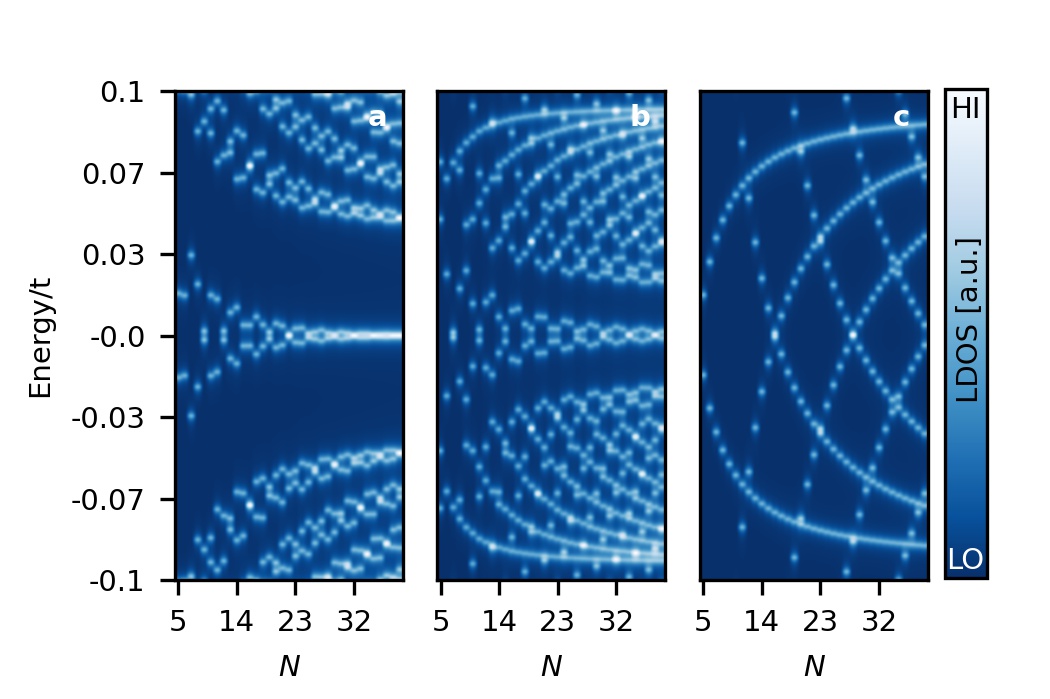} 
\caption{
\textbf{Low energy spectra vs.\ $N$ for model \eqref{toy} in the topologically nontrivial phase, OBC imposed.} \textbf{a,}  Large gap case, $(\Delta, \alpha, \mu, J)=(0.15, 0.1, 1.35, 6)t$. The MZMs do not oscillate because the bands are well-gapped so the coherence length is short and the MZMs do not hybridize. The MZMs are local to the ends of the chain; the coherence length is $\xi=5$. \textbf{b,}  Medium gap case, $(\Delta, \alpha, \mu, J)=(0.15, 0.1, 1.4, 3.5)t$. The MZMs hybridize weakly and oscillate in energy; the coherence length is $\xi=42$. \textbf{c,}  Small gap case, $(\Delta, \alpha, \mu, J)=(0.05, 0.01, 2.05, 3.6)t$. The hybridization is strong and the MZMs also completely intersect bulk states. This example shows a remarkable similarity with the experiment (Fig.\,4\,b of the main text); the coherence length is $\xi=170$.}
\label{fig:toy_model_spectra}
\end{figure}

\subsection{Coherence length}\label{coherence-length} 

As with the 80-band model we compute the approximate coherence length
\(\xi \sim v_F/\Delta_{\rm gap}\) (Supplementary Fig.\,\ref{fig:phasediagram}, third row) for the simplified model \eqref{toy} where $\Delta_{\rm gap}$ is the effective gap size shown in the second row of Supplementary Fig.\,\ref{fig:phasediagram}. Due to the small gap sizes, some of the coherence lengths are as large as 10\,000 or even 100\,000 sites. As a consequence, hybridization effects can be drastic. We can readily
see this in the zero-energy wavefunctions for open boundary conditions (OBC); examples are presented in Supplementary Fig.\,\ref{fig:coh_len_wf} for $N=151$ where for decreasing gap size the hybridization becomes more and more prominent.

\subsection{Energy spectra vs.\ chain length}\label{spectra} 

When the gap is large the coherence length is short and the MZMs are
well localised to the ends of the chain. Correspondingly, they are fixed
to zero-energy (Supplementary Fig.\,\ref{fig:toy_model_spectra}\,a). When the gap is
small, the coherence length is long and the MZMs hybridize, leading to the oscillatory pattern of the low-energy LDOS along the chain. Correspondingly, they split in energy and also oscillate as
a function of chain length (Supplementary Fig.\,\ref{fig:toy_model_spectra}\,b,c)\,\cite{stanescu-13prb094518,schneider2}.  
By comparison with the experimental Fig.\,4\,b in the main text, we can safely conclude that the experimental system corresponds to the ``small gap'' scenario.

\subsection{Chain embedded in extended substrate}\label{extended-substrate} 

\begin{figure}[t!]
\centering
\includegraphics[width=\linewidth]{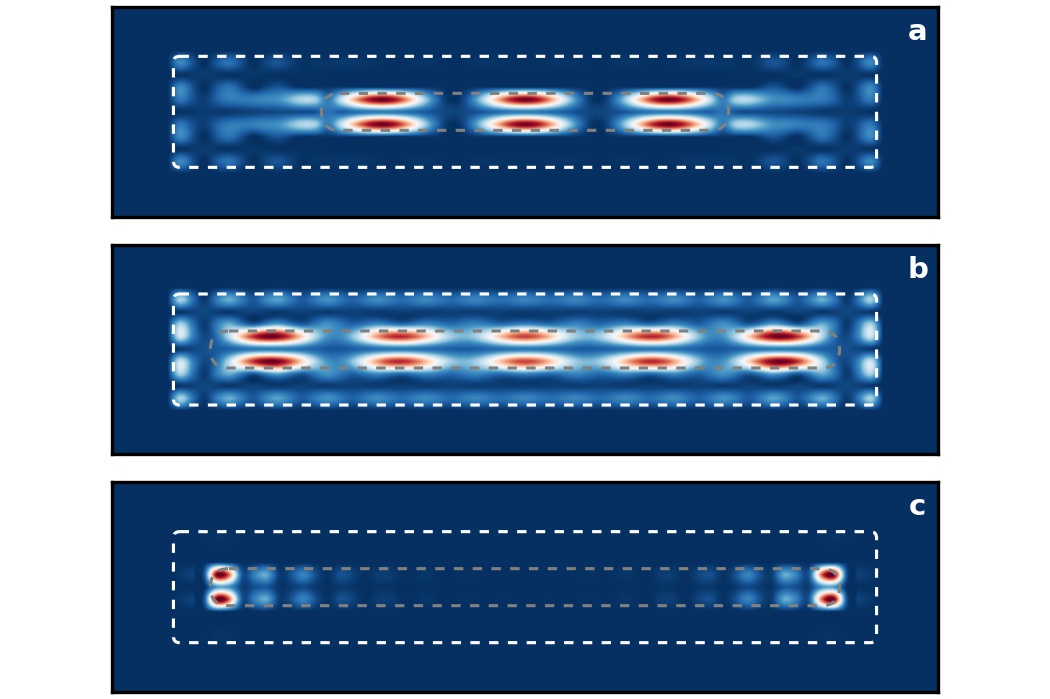}
\caption{\textbf{Zero-energy LDOS for the simplified model on an extended substrate.} The white dashed box indicates the size of the substrate while the grey dashed line marks the size of the chain (including the atoms next to it). \textbf{a,} $N=33$ chain on a substrate consisting of $57\times 3$ unit cells. The LDOS is very similar to Fig.\,3\,d of the main text. Parameters used $(\Delta, \alpha, \mu, J)=(0.05, 0.01, 2.05, 3.8)t$. \textbf{b,}  $N=51$ chain on a $57\times 3$ unit cell substrate. The LDOS is very similar to Fig.\,3\,e of the main text. Parameters used $(\Delta, \alpha, \mu, J)=(0.05, 0.01, 2.05, 3.95)t$. \textbf{c,}  $N=51$ chain on a $57\times 3$ unit cell substrate. The LDOS is very similar to Fig.\,3\,g of the main text. Parameters used $(\Delta, \alpha, \mu, J)=(0.05, 0.01, 1.35, 6)t$. OBCs are imposed in all subfigures.}
\label{fig:figS12}
\end{figure}

The results summarized in Fig.\,3 of the main text are based on a four-atomic unit cell. The observed side features are concentrated at the sides of the unit cell which is surrounded by hard-wall boundaries. 
In order to rule out confinement effects as the source of the observed side features, we have extended the simplified model, introduced in Fig.\,3 of the main text and further discussed previously, and increased the substrate size.
Essentially we have added more unit cells around the chain with the magnetic atom omitted. In addition, we scaled all the new hoppings by a global factor in order to slightly reduce them. As the DFT results have revealed, underneath the magnetic adatom, atomic positions are slightly contracted leading to a closer distance between the atoms. As a lowest order effect, we assume that the closer distance between two atoms results in a slightly increased hopping amplitude. It seems, hence, appropriate to slightly decrease the hoppings which are not within the unit cell that contains the magnetic atom. We have also varied the size of the substrate and considered asymmetric scenarios where the chain is not in the center but slightly off the center of the substrate. Moreover, we have studied a substrate with 3 and a half unit cells in the $y$-direction, but also situations with periodic boundary conditions imposed.

The results shown in Supplementary Fig.\,\ref{fig:figS12} summarize our analysis and essentially reproduce what is shown in Fig.\,3 in the main text. There are sometimes small differences and some parameter fine tuning might be necessary, but the important finding is that side features always persist and are thus not due to confinement. These results further substantiate our previous results.

%
%

\begin{figure}[t!]
\centering
\includegraphics[width=0.99\columnwidth]{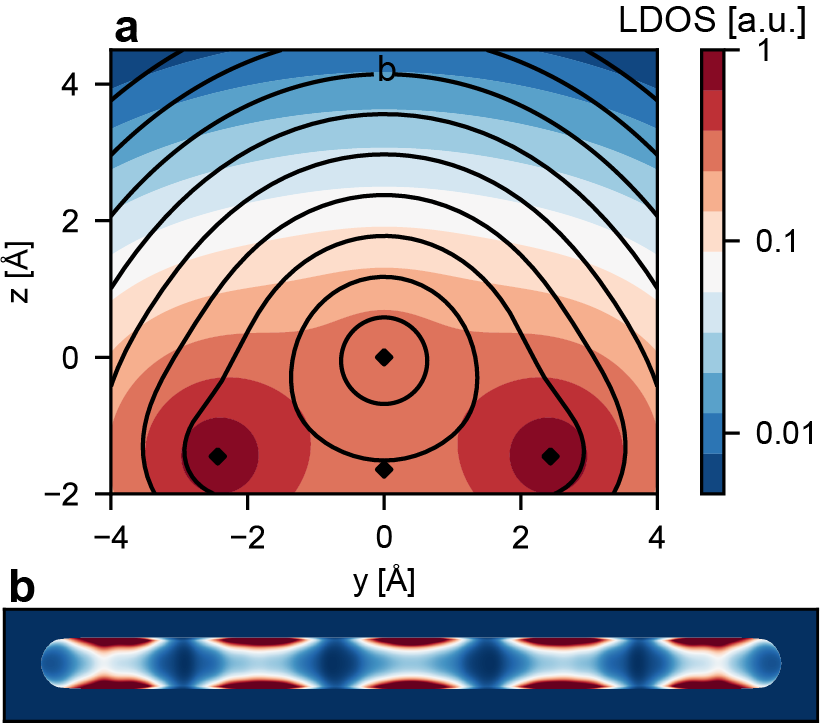} 
\caption{\textbf{Constant-current tip trajectory and zero energy LDOS}.
\textbf{a,} Cross-section through the middle of a Mn$_{51}$ chain from Fig.\,2\,\textbf{c} of the main text.
Constant current contours are shown as black lines.
The zero energy LDOS is shown as filled color contours (log scale).
Atom sites are shown as black diamonds.
\textbf{b,} $dI/dV$ map (linear scale) for the constant current setpoint labeled in \textbf{a}.
Parameters: $V_{\mathrm{stab}} = -30$ meV, $\xi=1$ \AA.} 
\label{fig:3DLDOS}
\end{figure}

\section*{Supplementary Note 4. Side-features in a constant-current feedback loop}\label{secA1}

In the main text, we show the LDOS evaluated in the plane of the chain.
However, as explained in the Methods section, the height of the tip apex is controlled by stabilizing the tip at a bias voltage $V_{\mathrm{stab}}$ and current $I_{\mathrm{stab}}$.
To compare the theoretical LDOS to the experimental $dI/dV$, we follow a phenomenological description of the tunneling given in Ref.\,\cite{feldman-17np286}:  
\begin{align}
    \frac{dI(\vec{r}, V)}{dV} \propto& \sum_i w(|\vec{r} - \vec{r}_i|) N(\vec{r}_i, eV), \\
    I(\vec{r}, V) \propto& \sum_i w(|\vec{r} - \vec{r}_i|) \int_0^{eV} d\omega N(\vec{r}_i, \omega), \\[5pt]
    w(r) =& \exp \left(-r/\xi\right).
\end{align}
The differential conductance, $dI(\vec{r}, V)/dV$, is a weighted sum of the LDOS, $N(\vec{r}_i, eV)$, weighted by an exponential decay, $w(r)$, depending on the distance between the tip and site $i$ and a decay length $\xi$.
Similarly, the current, $I(\vec{r}, V)$, is a weighted sum of the integrated LDOS with the same weight.

In Supplementary Fig.\,\ref{fig:3DLDOS}\,\textbf{a}, we show a cross section through the middle of a Mn$_{51}$ chain with the same parameters as Fig.\,2\,\textbf{c} in the main text.
As we follow a constant current line above a Nb site to the Mn site ($y=2.4$ \AA\ to $y=0$), the zero energy LDOS goes from a large intensity to a low intensity, resulting in a side feature.
In Supplementary Fig.\,\ref{fig:3DLDOS}\,\textbf{b}, we show an example of a $dI/dV$ map using a constant-current feedback loop.
Supplementary Fig.\,\ref{fig:3DLDOS}\,\textbf{b} has the same side-features as Fig.\,2\,\textbf{c} of the main text.

\section*{Supplementary Note 5. Extended spectroscopic data of sub-gap states around manganese and iron chains}

The experimental data in the main text only show the spatial distributions of selected states close to zero energy. In  Supplementary Fig.\,\ref{fig:exp1supp}, we show additional $dI/dV$ maps measured around Mn (panels a,b) and Fe (panel c) chains at different energies. It is clearly visible that the respective lowest-energy states (at $|E| < 150\,\mu$eV) are all located on the sides of the chain, in agreement with the conclusions of the main text. In contrast, the higher-energy excitations can also be located on the chain’s center as they belong to a band emerging from the hybridizing YSR states induced by a different orbital of Mn\,\cite{schneider-21np}.  

Notably, the LDOS at exactly zero energy around the two Mn chains shown in this figure is dominated by a defect-induced localized sub-gap state. The positions of the defects are marked by the green arrows in the constant-current images. At this position along the chain, a strong, near-zero energy state appears on the side of the chain. It is very likely that this state is not a property of the Mn chain itself but rather induced by the defect, which is why we chose to show the next higher energy excitation in the main text. A comparison of the datasets in Supplementary Fig.\,\ref{fig:exp1supp} with the zero-energy $dI/dV$ map in Fig.\,3 of Ref.\,\onlinecite{schneider-21np}    
shows that the side features are typically distributed along the entire side of the chain (cf.\ Fig.\,3 in Ref.\,\onlinecite{schneider-21np}   
with Fig.\,2\,c and Fig.\,3\,d of the main text of the present work). Thus, it is highly likely that the zero energy states ($E=0\,\mu$eV) localized near the defect in Supplementary Fig.\,\ref{fig:exp1supp}\,a,b are induced by this local perturbation. Moreover, we emphasize that the defect could possibly act as a scattering center for the interfering Bogoliubov-de-Gennes quasiparticles\,\cite{schneider-21np}   
at all energies in the chain, thereby perturbing the oscillatory pattern of the sub-gap states.

\begin{figure*}
\centering
\includegraphics[width=1.99\columnwidth]{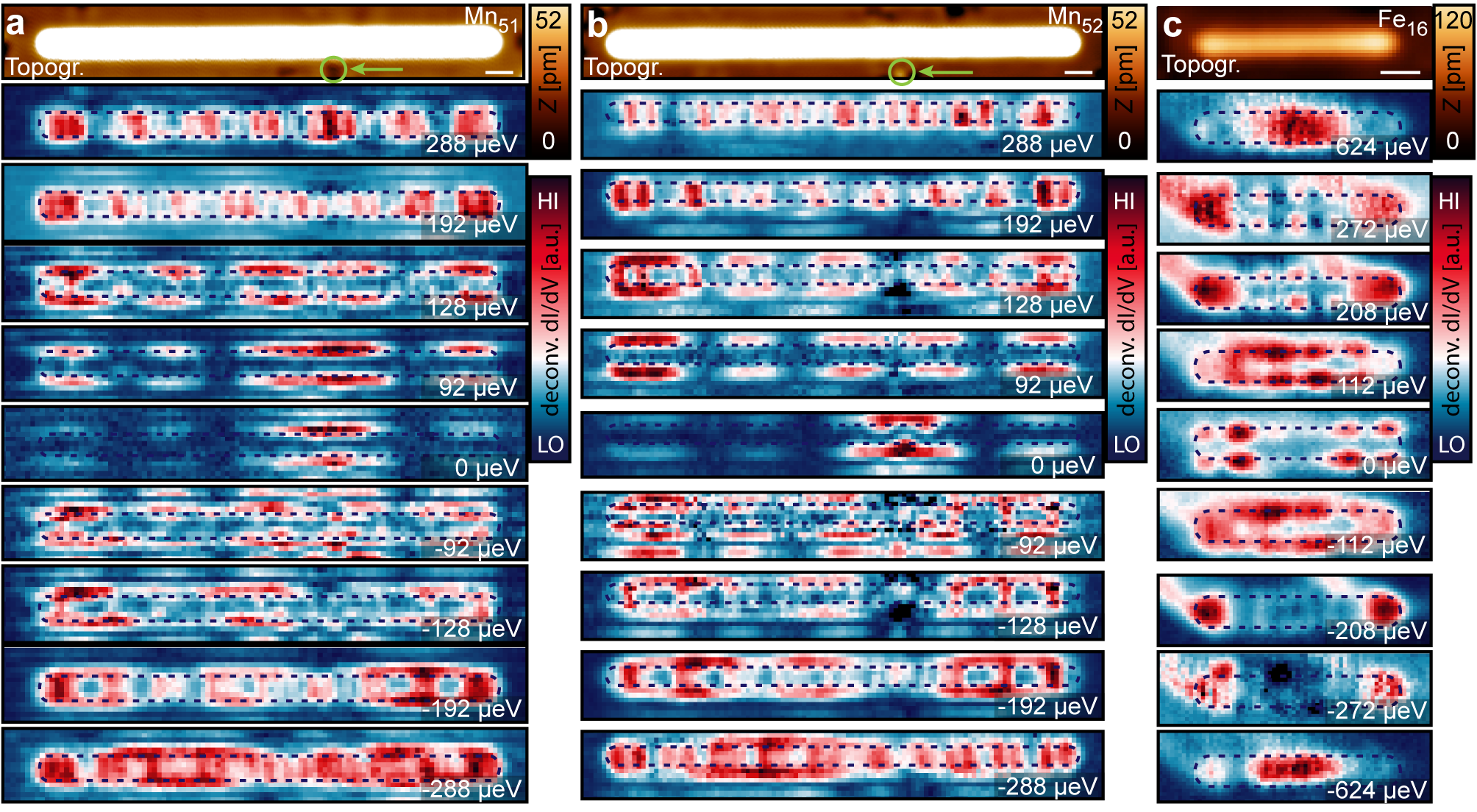}
\caption{\textbf{Extended spectroscopic measurements of sub-gap states around linear Mn and Fe chains. a,} Constant-current image with enhanced contrast (top panel) and deconvoluted $dI/dV$ maps (bottom panels) at various sub-gap energies as indicated on each panel for the Mn$_{51}$ chain presented in Fig.\,4\,a of the main text. The white bar corresponds to 1\,nm and the green arrow highlights the position of a defect on the surface inducing additional sub-gap states. \textbf{b,} Constant-current image (top panel) with enhanced contrast and deconvoluted $dI/dV$ maps (bottom panels) at various sub-gap energies for a Mn$_{52}$ chain along the [001] direction. The white bar corresponds to 1\,nm and the green arrow highlights the position of a defect on the surface inducing additional sub-gap states. \textbf{c,} Constant-current image (top panel) and deconvoluted $dI/dV$ maps (bottom panels) at various sub-gap energies for the Fe$_{16}$ chain presented in Fig.\,4\,c of the main text. The white bar corresponds to 1\,nm. The apparent extent of the chains from the constant-current images is marked by the dark dashed lines in all $dI/dV$ maps. Parameters: $V_{\rm stab} = -6$\,mV, $I_{\rm stab} = 1$\,nA, $V_{\rm mod} = 20\,\mu$V.
}
\label{fig:exp1supp}
\end{figure*}

\section*{Supplementary Note 6. Sub-gap states induced by iron atoms and dimers}

Single Mn atoms on Nb(110) have recently been found to induce pairs of spatially anisotropic multi-orbital YSR states\,\cite{schneider-21np,beck-21nc2040,kuester-21nc1108},   
which are oriented along the high-symmetry directions of the underlying substrate. Accordingly, the localization of sub-gap features along the sides of Mn chains on Nb(110) was explained by a band formed by the hybridization of YSR states stemming from a particular orbital featuring extended lobes along the [1$\bar{1}$0] direction\,\cite{schneider-21np}.   
In contrast, single Fe atoms on Nb(110) only show
resonances very close to the coherence peaks of Nb\,\cite{odobesko-20prb174504,kuester-21nc1108}    
at $E\approx \pm\Delta_{\rm Nb} = \pm 1.52$\,meV, indicating that the YSR states are almost merged with the coherence peaks. These states are shown in Supplementary Fig.\,\ref{fig:exp2supp}\,a, b and do not feature a distinct anisotropic spatial distribution. Therefore, the side localization of the observed features in Fe chains cannot be explained just by linear combination of the YSR states of the individual constituents. However, dimers of Fe atoms\,\cite{friedrich-21prb235437}     
constructed along the [001] direction (Supplementary Fig.\,\ref{fig:exp2supp}\,c) do already show pronounced YSR states at lower energies, i.e., approaching $E_F$. But, they are still mainly located on the center of the dimer and not on the sides. Therefore, the side features around Fe chains shown in Fig.\,4\,c of the main text are a surprising observation not obviously 
related to YSR physics of the individual atoms.

\begin{figure*}
\centering
\includegraphics[width=1.99\columnwidth]{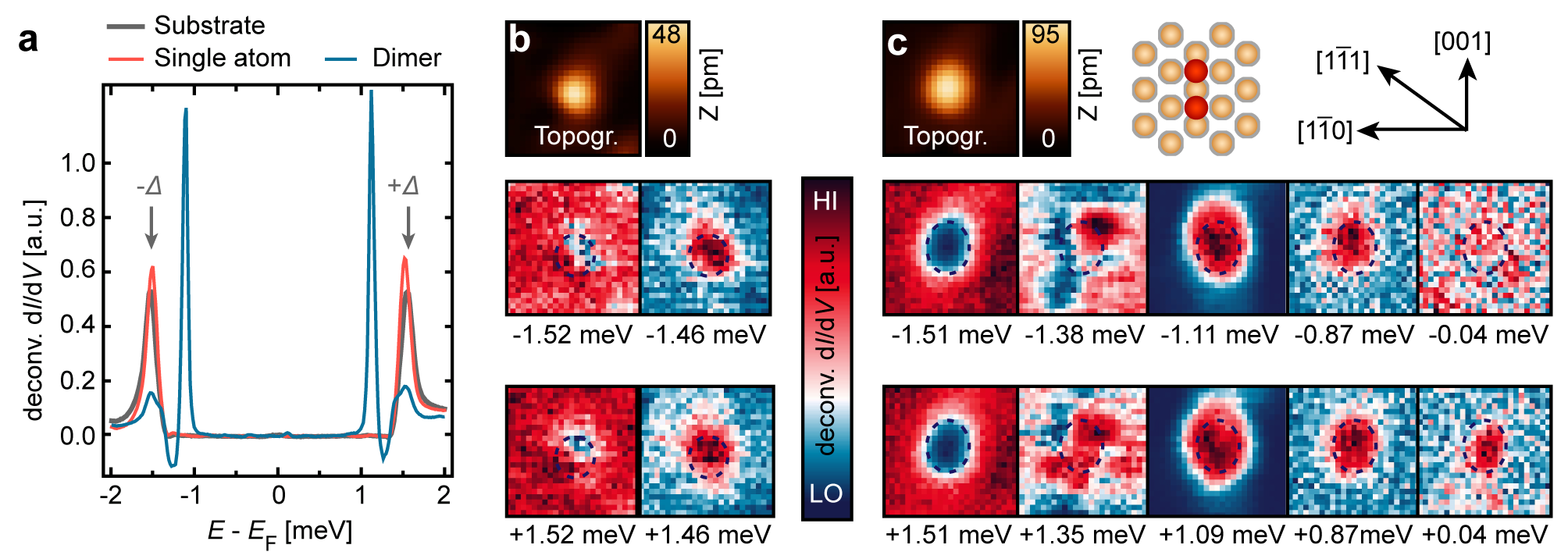}
\caption{\textbf{Sub-gap states induced by Fe atoms and dimers on Nb(110). a,} Deconvoluted $dI/dV$ spectra measured on the bare Nb substrate (gray), on a single Fe atom (red) and on an Fe dimer along the [001] direction (blue). The energetic position of the Nb substrate’s coherence peaks is marked by $\pm\Delta$. \textbf{b,} Constant-current image (top panel) and deconvoluted $dI/dV$ maps (bottom panels) at sub-gap energies as indicated in the proximity of a single Fe atom (image sizes are 2 x 2 nm$^2$). \textbf{c,} Constant-current image (top panel) and deconvoluted $dI/dV$ maps (bottom panels) at sub-gap energies around an Fe dimer along the [001] direction. The geometry of the dimer is sketched to the right of the constant-current image (brown spheres: Nb atoms, red spheres: Fe atoms, image sizes are 2.5 x 2.5 nm$^2$). The apparent extent of atom and dimer, respectively, are marked by the black dashed lines in the $dI/dV$ maps. Parameters: $V_{\rm stab} = -6$\,mV, $I_{\rm stab} = 1$\,nA, $V_{\rm mod} = 20\,\mu$V.
}
\label{fig:exp2supp}
\end{figure*}